\documentclass[preprint,aps,showkeys]{revtex4-1}

\usepackage{amsmath}
\usepackage{amssymb}
\usepackage{graphicx}
\usepackage{mathtools}
\usepackage{color}
\usepackage{fancyhdr}
\usepackage{txfonts}
\usepackage{upgreek}
\usepackage[greek,english]{babel}
\usepackage{setspace}

\newcommand{\deq}{\overset{\operatorname{def}}{=}}

\newcommand{\Ei}{\mathrm{Ei}}
\newcommand{\intl}[2]{\!\underset{#1}{\overset{#2}{\rotatebox[origin=rc]{15}{\large\ensuremath{\int}}}}}

\newcommand{\inth}[2]{\!\!\!\underset{#1}{\overset{#2}{\rotatebox[origin=rc]{15}{\huge\ensuremath{\int}}}}\!\!}
\newcommand{\D}{\partial}

\renewcommand{\alpha}{\alphaup}

\renewcommand{\beta}{\betaup}
\renewcommand{\gamma}{\gammaup}
\renewcommand{\delta}{\deltaup}

\renewcommand{\epsilon}{\varepsilonup}
\renewcommand{\varepsilon}{\epsilonup}

\renewcommand{\zeta}{\zetaup}

\renewcommand{\eta}{\etaup}
\renewcommand{\theta}{\thetaup}
\renewcommand{\vartheta}{\varthetaup}

\renewcommand{\iota}{\iotaup}

\renewcommand{\kappa}{\varkappa}
\renewcommand{\lambda}{\lambdaup}

\renewcommand{\mu}{\muup}

\renewcommand{\nu}{\nuup}
\renewcommand{\xi}{\xiup}

\renewcommand{\pi}{\piup}

\renewcommand{\rho}{\rhoup}
\renewcommand{\varrho}{\varrhoup}
\renewcommand{\sigma}{\sigmaup}
\renewcommand{\varsigma}{\varsigmaup}

\renewcommand{\tau}{\tauup}
\renewcommand{\Upsilon}{\textrm{\greektext U}}
\renewcommand{\upsilon}{\upsilonup}
\renewcommand{\phi}{\upvarphi}
\renewcommand{\varphi}{\phiup}

\renewcommand{\chi}{\chiup}
\renewcommand{\psi}{\textrm{\greektext y}}
\renewcommand{\omega}{\omegaup}

\begin{document}

\title{Selection strategies for randomly partitioned genetic replicators}
\thanks{The authors thank Marco Ribezzi-Crivellari, David Lacoste, Clement Nizak, Olivier Rivoire, and Zorana Zeravcic for fruitful discussions. This work was supported by European Research Council (ERC, Consolidator Grant 647275 ProFF).}

\author{Anton S. Zadorin}
\affiliation{Laboratoire Gulliver, CNRS, ESPCI Paris, PSL Research University}
\affiliation{Current affiliation: LBC, CBI, ESPCI Paris, PSL Research University}
\author{Yannick Rondelez}
\email{yannick.rondelez@espci.fr}
\affiliation{Laboratoire Gulliver, CNRS, ESPCI Paris, PSL Research University}

\begin{abstract}
The amplification cycle of many replicators (natural or artificial) involves the usage of a host compartment, inside of which the replicator express phenotypic compounds necessary to carry out its genetic replication. For example, viruses infect cells, where they express their own proteins and replicate. In this process, the host cell boundary limits the diffusion of the viral protein products, thereby ensuring that phenotypic compounds, such as proteins, promote the replication of the genes that encoded them. This role of maintaining spatial co-localization, also called genotype-phenotype linkage, is a critical function of compartments in natural selection. In most cases however, individual replicating elements do not distribute systematically among the hosts, but are randomly partitioned. Depending on the replicator-to-host ratio, more than one variant may thus occupy some compartments, blurring the genotype-phenotype linkage and affecting the effectiveness of natural selection. We derive selection equations for a variety of such random multiple occupancy situations, in particular considering the effect of replicator population polymorphism and internal replication dynamics. We conclude that the deleterious effect of random multiple occupancy on selection is relatively benign, and may even completely vanish is some specific cases. In addition, given that higher mean occupancy allows larger populations to be channeled through the selection process, and thus provide a better exploration of phenotypic diversity, we show that it may represent a valid strategy in both natural and technological cases.
\end{abstract}


\maketitle

\section*{Introduction}
Genetically encoded replicators, such as viruses, plasmids, or artificial DNA constructs, typically rely on an indirect replication strategy. For example, the RNA of a virus such as bacteriophage Q$\beta$ uses the 
translation machinery of its host bacterial cell to first express a replicase, which accumulates in the cellular cytoplasm \cite{Manrubia:2006viral}. At the 
second step, this catalytic protein picks up its own encoding mRNA and generates copies of it, thereby closing the replication loop of the viral 
genome. The phenotypic intermediate compound need not be directly involved in genetic amplification. For example, the resistance proteins expressed by 
a plasmid participate in the spreading of the genetic element by promoting the selective survival of its host cells. In fact, most (but not all 
\cite{Flores:2004eh}) parasitic replicators use a similar \emph{trans} strategy: their genetic information codes for diffusible compounds that are 
first expressed in many copies inside the host, and then act globally to promote the propagation of the genetic carrier.

Beside biological parasitic replicators, a number of recent works have  attempted to mimic natural evolution and to create artificial molecular 
replication loops for technological applications. These approaches create indirect replication strategies \cite{Mastrobattista:2005hig,Esvelt:2011cv,Davidson:2012gma,Ellefson:2014hi,Skirgaila:2013comp}, and also use compartments. For example, Holliger and coworkers designed an experimental multistep replication protocol (which includes bacterial transformation, emulsification, thermocycling, and recloning), by which plasmids encoding for active PCR (Polymerase Chain Reaction) enzymes can propagate \cite{Ghadessy:2001hj}. Another approach called genetic complementation is more closely related to the natural plasmid replication strategy: one first creates a cell that is deficient in a molecular function necessary for its growth. Exogenous plasmids encoding for products able to restore the deficient function are then introduced in the cell and are selectively propagated by the survival and growth of the "complemented" hosts \cite{Alcalde:2017}. 
\begin{figure}[tbh]
\centering
\includegraphics[width=0.8\textwidth]{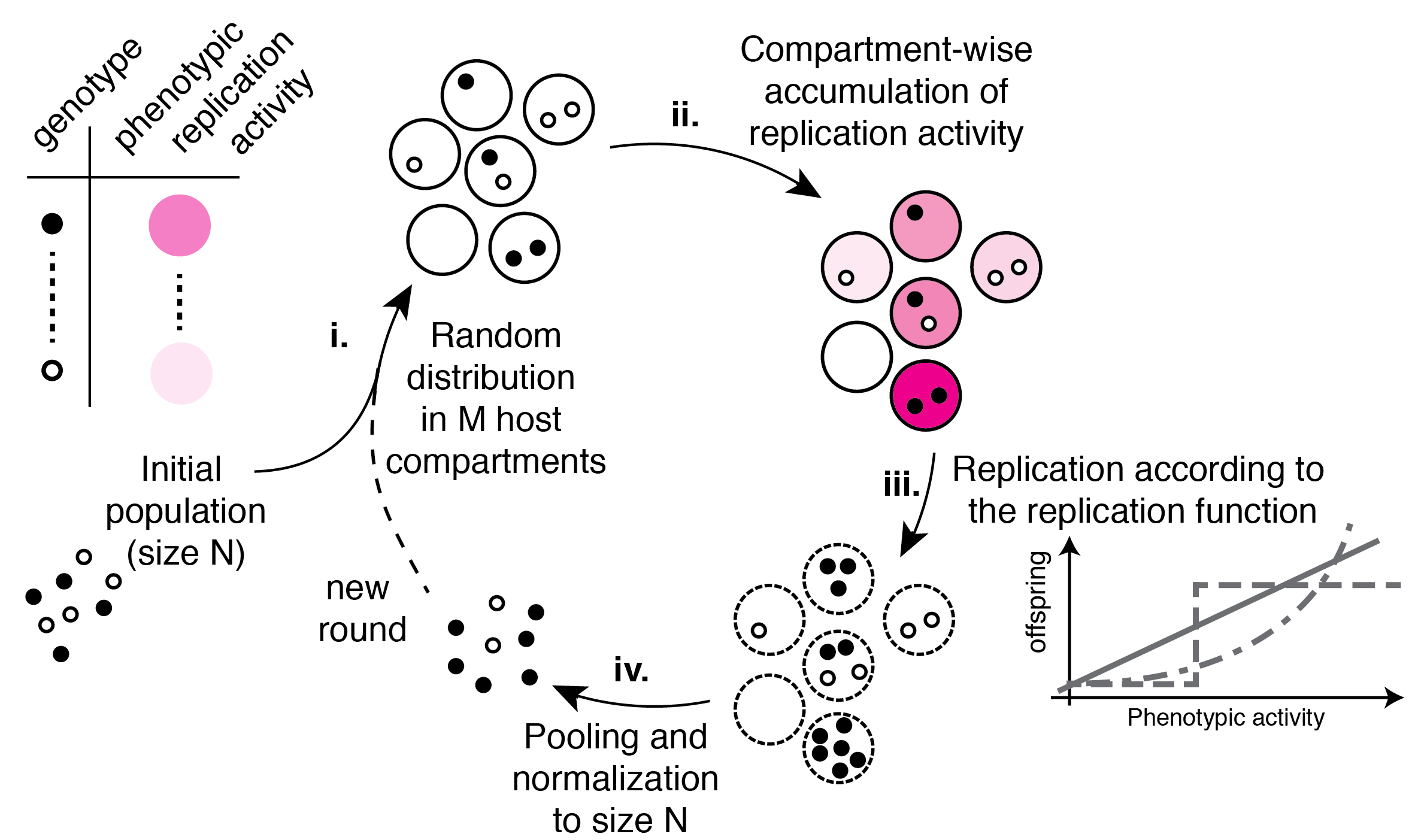}

\caption{Selection of replicators with random partitioning. i. Genetic replicators are randomly distributed among available host compartments. ii. In each compartment, genetic replicators express trans-acting compounds required to complete their replication cycle, i.e. generate a local replication activity. Some compartments contain multiple replicators, with possibly different phenotypes, collectively contributing to the global activity in the compartments. iii. The activity in a given compartment applies equally to all local variants, irrespective of their individual contribution. The number of generated offspring (copies) is linked to the total activity via the 'replication function', which may take different shapes. iv. The compartments are broken and all replicators are pooled before a new cycle is started, at constant population size. Random co-encapsulation, representative of many real-world situations, creates hitch-hiking opportunities that benefit weak phenotypes and are thus expected to negatively affect the efficiency of the natural selection loop.}

\label{fig:1}
\end{figure}

Because the intermediate compounds are diffusible, physical encapsulation is essential both to maintain the genotype-phenotype proximity linkage and to achieve reasonable local concentrations from a single genetic copy. For natural parasitic replicators, such as viruses or plasmids, containment is provided by the host cell. The products expressed by the viral genome accumulate in the infected cell, in direct proximity to their encoding genetic polymer. Artificial \emph{in vitro} replicators depend on man-made compartments, for example emulsions \cite{Mair:2017}.

Importantly, invasion of the host compartments (\emph{via} infection, transformation, encapsulation\ldots) is usually a random process governed by the Poisson statistics. The case of exclusively single occupancy is therefore only relevant at extremely low ratios of replicators to hosts. When significant populations of replicators are considered, co-occupancy of one host by multiple replicators becomes commonplace. Random multiple infection is an important feature of viral or phages life cycles \cite{Frank:2001,Novella:2004ec,Manrubia:2006viral}, underlying for example the phenomenon called "multiplicity reactivation" in influenza \cite{Farrell2019}.  For plasmids, transformation of E. coli cells is known to lead to cells containing multiple plasmids with a negative effect on selections and screens \cite{goldsmith2007avoiding}. Random partitioning also affects eukaryotic mitochondrial DNA populations during early development, where it is thought to exert a purifying effect against defective copies \cite{johnston2015stochastic}.

In the context of a population of genetically diverse replicators, a host compartment may thus contain variants expressing different phenotypes. Inside the host compartment, a global replication activity is generated and applies indiscriminately to all local variant genomes. Therefore, each replicator experiences a replication phenotype that is integrated over the co-encapsulated variants. This pooling of phenotypes benefits the weakest replicator \cite{Novella:2004ec}, and may also penalise the strongest (Fig~\ref{fig:1}). One intuitively feels that high occupancy should slow down natural selection by interfering with the spontaneous enrichment of the most efficient replicators from the population  \cite{Baret:2009}. In the extreme case were a single compartment is available for all replicators, it is clear that the activity-averaging effect would even completely hinder any progress of the population. 

In this paper, we quantify the selection slow-down effect associated with random (Poissonian) compartmentalization process under assumptions relevant to both viral replication and artificial molecular evolution protocols. In particular, we assume additivity of the phenotypic activities and consider different possible shape of the \emph{replication functions} $f$, i.e. the function that links the total activity in a given compartment to the number of offspring it will eventually produce. In principle, depending on the physical chemistry of the replication/selection cycle, linear, nonlinear, saturating or even discontinuous functions may be considered \cite{Alcalde:2017,Rosinski:2002}.

\section*{Materials and Methods}
\label{matmet}

\subsection*{Derivation of the update equations for bivariate phenotypic distributions}

Let each genotype be characterised by a phenotypical trait $x$, which we will refer to as \emph{activity}. It may be a replicative activity of 
individuals themselves or some other enzymatic activity that an experimetalist is interested by and thus selects for. Consider a population of 
replicator with only two genotypes, with phenotypic activities $x_1 > x_2$ (an improved mutant and the wild type) and gene frequency $p$ and $1 - p$. 
As these two frequencies are not independent, we can consider only genotype~1 and its frequency $p$ in the following.

The \emph{fitness}, $w_1$, of genotype~1 is the average number of descendants left by replicators with genotype~1 after the selection step. Likewise, 
$w_2$ will denote the fitness of genotype~2. The \emph{mean fitness} $\bar w$ is given by $pw_1 + (1-p)w_2$ and is basically equal to the overall 
growth rate of the whole population. The gene frequency of genotype~1 at the next round, $p'$, is then given by the formula
	\begin{equation}
	p' = \frac{w_1}{\bar w}\,p.
	\end{equation}

\noindent The change in gene frequency $\Delta p = p' - p$ is then given by the well known formula \cite[page 9]{Nagylaki:1992int}:
	\begin{equation}
	\Delta p = p(1-p)\frac{w_1 - w_2}{pw_1 + (1-p)w_2}.
	\label{update-equation}
	\end{equation}

The explicit form of this \emph{update equation} for different cases is our main goal. To find it, we need to compute the fitnesses $w_i$.

Let us consider a compartment with $i$ replicators of genotypes~1 and $j$ replicators of genotypes~2 (\emph{$ij$-compartment}). An essential assumption is that \emph{each} replicator in that particular compartment generates the same number of offspring as a result of selection, regardless of its identity. This number $w_{ij}$ depends only on the overall activity in the compartment and thus only on numbers $i$ and $j$ (and, of course, on $x_1$ with $x_2$). The number $w_{ij}$ can be called the \emph{local fitness} in an $ij$-compartment.

We will denote the numerical ratio of replicators to host compartments, or \emph{mean occupancy}, by $\lambda$ (Fig~\ref{fig:2}A). We assume the Poisson law of compartmentalization with the probability to find a compartment with $n$ individuals (of whatever phenotype)
	\begin{equation}
	P_n = \frac{e_{}^{-\lambda}\lambda^n}{n!},
	\end{equation}

\noindent and with no preference of one phenotype over the other to be included into a compartment. The probability $P_{ij}$ to find an $ij$-compartment is then given by the combination of the Poisson statistics and the binomial distribution:
	\begin{equation}
	P_{ij} = \frac{e_{}^{-\lambda}\lambda^{i+j}}{(i+j)!} C^i_{i+j} p^i (1-p)^j = \frac{e_{}^{-\mu}\mu^i}{i!}\frac{e_{}^{-\nu}\nu^j}{j!},
	\end{equation}

\noindent where $C^k_n = n!/(k!\,(n-k)!)$ is the binomial coefficient, $\mu = p\lambda$, and $\nu = (1-p)\lambda$.

By definition, fitness $w_1$ is equal to the ratio of the total number of offspring (after selection) of all replicators with genotype~1 to the initial number of these replicators. In the considered limit it is given by the following expression:
	\begin{equation}
	w_1 = \frac{\sum\limits^{\infty}_{i=1}\sum\limits^{\infty}_{j=0} P_{ij}i w_{ij}}
	{\sum\limits^{\infty}_{i=1}\sum\limits^{\infty}_{j=0} P_{ij}i}.
	\end{equation}

Likewise, fitness of phenotype~2 is equal to
	\begin{equation}
	w_2 = \frac{\sum\limits^{\infty}_{i=0}\sum\limits^{\infty}_{j=1} P_{ij}j w_{ij}}
	{\sum\limits^{\infty}_{i=0}\sum\limits^{\infty}_{j=1} P_{ij}j}.
	\end{equation}

These expressions of $w_i$ allow to rewrite the update equation (\ref{update-equation}) in terms of activities $x_i$, given the dependence of $w_{ij}$ on the activities. However, sometimes it is more straightforward to compute the difference $w_1 - w_2$ and the mean fitness $\bar w$ directly.

We will consider different models for $w_{ij}$. All of them essentially assume either the form
	\begin{equation}
	w_{ij} = \frac{f(ix_1 + jx_2)}{i+j}
	\label{sharing}
	\end{equation}

\noindent or the form
	\begin{equation}
	w_{ij} = f(ix_1 + jx_2),
	\label{non-sharing} 
	\end{equation}

\noindent where $f$ is the \emph{replication function} that defines the full replication activity of the compartment. We refer to situation (\ref{sharing}) as \emph{sharing} because the offspring possibilities are divided equally between local replicators. In situation (\ref{non-sharing}), called \emph{non-sharing}, the full replication activity applies  to each replicator. For example, if a compartment has local activity for 100 replications, and two replicators are present, each will be copied 50 times in the sharing case, and each will be copied 100 times in the non-sharing case, with final total population 100 or 200, respectively.  These categories are appropriate in situations where the phenotypic activity is stoichiometric or catalytic, respectively. The detailed derivations of the update equations for the linear replication function and for the step replication function with and without sharing can be found in Appendices~\ref{appA}--\ref{appD}.
 
More complicated nonlinear replication functions and polymorphic populations are treated with more general but more abstract mathematical methods described in our previous work \cite{Zadorin:2017} (see Appendices \ref{appA}, \ref{appB}, \ref{appE}, and \ref{appF}).

\subsection*{Numerical simulations}

The simulations were performed with Wolfram Mathematica. The number of compartments ($M$) was fixed in each trajectory. An initial set of $N = \lfloor 
\lambda M \rfloor$ activities was then drawn either from a bivariate distribution or from the corresponding continuous distribution.

One generation is implemented using the following loop, which is then repeated an appropriate number of times to get the needed trajectory:
\begin{itemize}

\item Each value from the set is randomly assigned to a compartment.

\item Each compartment is given a fitness in accordance the total internal activity and the replication function.

\item An updated weight for each activity value is obtained by summing the fitness of each of its representative in each compartment (i.e. the number of 
offspring it was able to create overall, taking into account the sharing behavior).

\item A new set of $N$ phenotype values is drawn randomly from the list of activities, using the 
updated weight list.

\end{itemize}

\vspace{1cm}
The authors affirm that all data necessary for confirming the conclusions of the article are present within the article, figures, and tables.

\section*{Results}
\subsection*{Resilience of selection to random distribution in compartments}
We start our analysis with a simple case where the replicator population contains $N$ individuals of only two genotypes, with phenotypic activities $x_1 > x_2$ (an improved mutant and the wild type) and frequency $p$ and $1 - p$. The replication function that we consider here is linear such that, after rescaling, $f(x) = x$. We also assume that the local replication activity inside the host is partitioned in equal proportion to each local replicators (the case of shared offspring possibility). 

The application of the computation method outlined in the Materials and Methods section to this case results in the following update equation (see details in Appendix~\ref{appA}, Section~1):
	\begin{equation}
	\Delta p = g(\lambda)\frac{p(1-p)(x_1 - x_2)}{px_1 + (1-p)x_2},\quad g(\lambda) \deq \frac{1 - e_{}^{-\lambda}}{\lambda}.
	\label{g-linear-update}
	\end{equation}

\begin{figure*}[tbh]
\centering
\includegraphics[width=.8\textwidth]{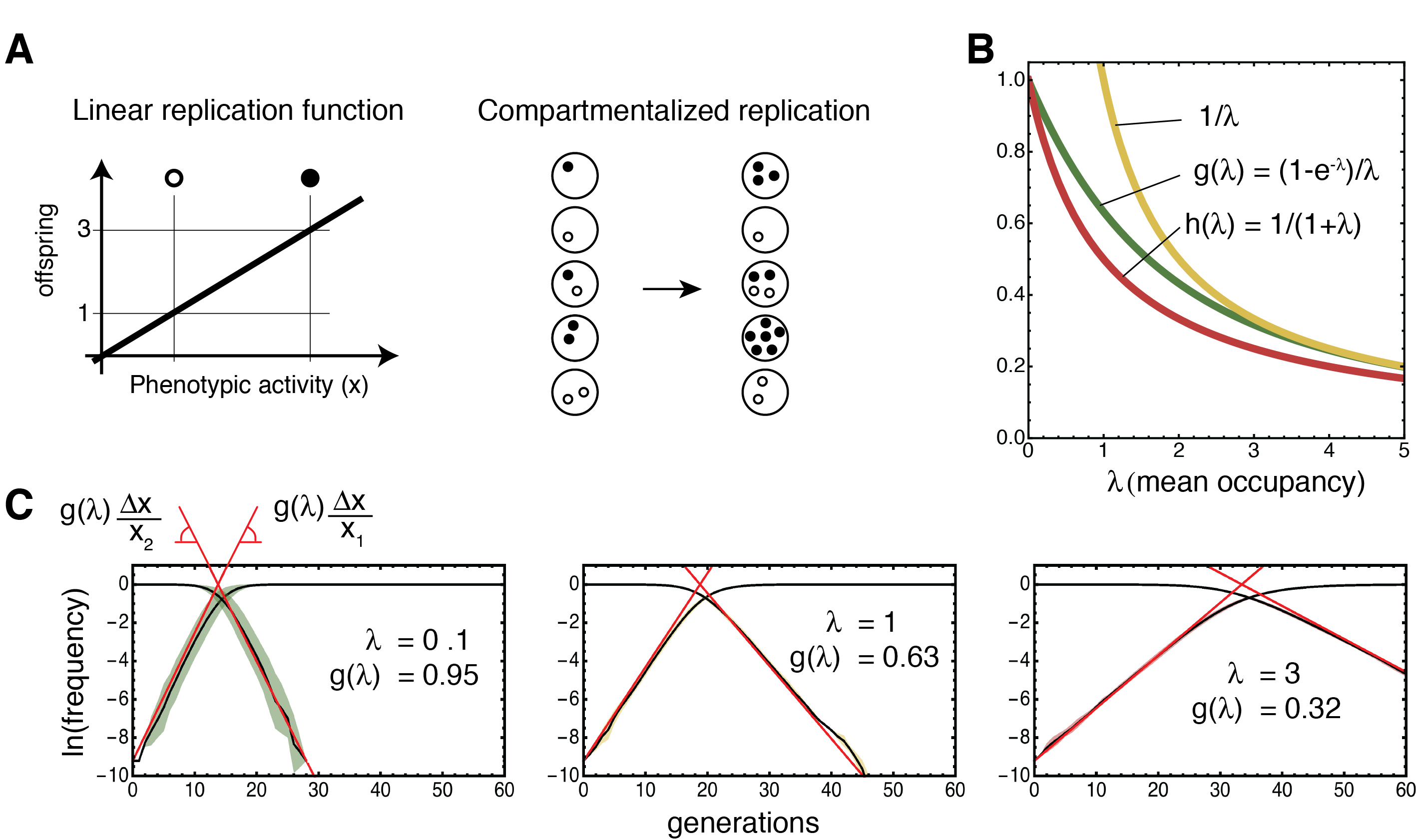}

\caption{{\bf A.}~In the linear replication case, the number of offspring is proportional to the phenotypic activity. The scheme shows an exemplary replication cycle for randomly compartmentalized replicators with this replication function, in the sharing case {\bf B.}~The function $g(\lambda)$ encapsulates the deleterious effect of random partitioning on natural selection in the linear and sharing case. The function $h(\lambda)$ is valid in the linear and non-sharing case. Both have $1/\lambda$ as asymptote at high load. {\bf C.} Simulated selection dynamics for the replacement of a wild type phenotype $x_2$ by a single fitter variant $x_1$, at various $\lambda$ and with $10^5$ hosts and $p_\mathrm{ini} = 10^{-5}$. Evolution of the frequency $p$ of $x_1$ along generations. The mean (black curves) and s.d. of 10 numerical runs is represented twice on each graph as $\ln(p)$ and $\ln(1-p)$ to show the asymptotic behaviour at small and large time. The classical time scales associated with invasion ($\Delta x/x_1$) and fixation ($\Delta x/x_2$) of the beneficial variant are now modulated by $g(\lambda)$ (red curves), and $t_\mathrm{tot} = t_1 + t_2$ provides a typical total time of a selection sweep by a single beneficial mutation. }

\label{fig:2} 
\end{figure*}

In the limit $\lambda \to 0$, where each mutant has its own compartment, $g(\lambda) \to 1$ and we recover the classical selection model, where the activity is understood as fitness \cite{Nagylaki:1992int}. Random partitioning is thus represented by a simple fitness-blurring factor $g(\lambda)$ that depends only on the mean occupancy $\lambda$. Enrichment by natural selection is controlled by this factor and decreases with $\lambda$ (Fig~\ref{fig:2}B).  This confirms that random partitioning and multiple occupancy indeed slow down the selection process, which `sees' an activity difference modulated by $g(\lambda)<1$.

When $p$ is very small, we can linearize (\ref{g-linear-update}), giving:

	\begin{equation}
	p' = \left(1 + g(\lambda)\frac{x_1 - x_2}{x_2}\right)p = \alpha p.
	\end{equation}

\noindent The population dynamic then follows as $p_t = \alpha^t p_0$, where $t$ is the number of generations. Similarly, the solution near $p = 1$ can be approximated by $1 - p_t = \beta^t(1 - p_0)$, with $\beta = 1 - g(\lambda)(x_1 - x_2)/{x_1}$. Note that the ratio $s = (x_1 - x_2)/x_2$ is traditionally called the selection coefficient. The effect of increasing $\lambda$ at low $p$ can thus be understood as the reduction of the selection coefficient by $g(\lambda)$.

If the real population size is $N$, a new improved mutant will have $p_0 = 1/N$. The conditions $\alpha^{t_1} p_0 = 1$ and $\beta^{-t_2}q_f = 1$, (where $q_f = 1 - p_f = 1/N$) provide estimates for the timescales associated with respectively, the invasion of the population by a new phenotype, and its fixation:
	\begin{equation}
	t_1 = \frac{\ln N}{\ln \alpha}\quad \text{and}\quad t_2 = -\frac{\ln N}{\ln \beta}.
	\label{t1}
	\end{equation}

Fig~\ref{fig:2}C provides a visual illustration of the  effect of random partitioning on selection dynamics. If population ratio changes are small at each round, this analysis can be extended to the continuous case, with similar conclusions (see Appendix~\ref{appA}, Section~2).

Two hypotheses can be invoked to explain the slowdown effect due to multiple occupancy. The first mechanism is a blurring of the phenotype-genotype linkage: good variants now obtain on average a lower replication efficiency, while bad variants get on average a higher one. The second mechanism could be the sharing of the replication ability: good variants get less offspring, because they have to share some of their replication activity with co-compartmentalized, less fit, variants. However, one can show that, if the total activity in one compartment is not shared but is instead fully assigned to each local replicator, all previously derived results hold, except that the fitness-blurring function $g(\lambda)$ is replaced by $h(\lambda)=1/(1+\lambda)$ (see Appendix~\ref{appB}). As $h<g$ for all $\lambda > 0$ (Fig~\ref{fig:2}B), we obtain the counterintuitive result that sharing is not a cause of the slowdown, but on the contrary tends to rescue the effect of fitness blurring. This surprising effect can be understood if one considers that sharing actually decreases the relative contribution of multiply occupied hosts in the composition of the next generation, with respect to less occupied hosts. Compared to the non-sharing situation, sharing is therefore closer to the case of exclusive single infection. More generally, in our infinite population model, any mechanism that would lower the total output of multiply infected hosts would tend to alleviate the blurring effect of multiple occupancy.

The main conclusion, valid in the linear selection case, is that multiple random occupancy affects the enrichment process through the fitness-blurring $g$ function, which can be regarded as imposing a relatively benign slowdown of the selection dynamics. For example going from $\lambda = 0.01$ to $\lambda = 1$, the fraction of hosts encapsulating more than one genome changes from 0.005\% to 26\%, but the rate at which the improved phenotype invades the population decreases by just 36\%. Importantly this same increase in $\lambda$ also implies a 100-fold jump in the replicator population size, greatly increasing the number of genetic variants that can be scanned by the selection process.

\subsection*{Step replication function}
The number of offspring generated does not need to linearly follow the phenotypic activity under selection. For example, in a technological context, beneficial mutants are often selected through a process called screening \cite{Mair:2017}, where every host compartment is individually observed and only those where activity is measured above a given threshold are picked up and sent to the next generation (all other are simply discarded, Fig~\ref{fig:3}A). If the threshold is set as close as possible to the phenotype of the highest activity, $x_1$ (see Appendix~\ref{appC} for the case of a different threshold), this process can be described by a step replication function $f(x) = \theta(x - x_1)$. We will use here the non-sharing case, which better describes a typical screening protocol; the sharing case is given in Appendix~\ref{appD}. For two variants $x_1$ and $x_2$ as above, one can show that frequency change in one round then goes as:
	\begin{equation}
	\Delta p = \frac{p(1-p) e_{m-2}\big(\lambda(1-p)\big)}{e^\lambda - (1-p) e_{m-2}\big(\lambda(1-p)\big)},
	\label{update-cutoff}
	\end{equation}

\noindent where $e_n$ is the truncated Taylor series of the exponential function to the $n$-th term and $m = \lceil x_1/x_2 \rceil$, that is, the minimum number of wild type mutants that one has to pack in a host to reach the selection threshold. The most interesting case (the emergence of rare beneficial mutants in the population) happens at small $p$, where linearization gives:

	\begin{equation}
	p' = \frac{1}{1 - e^{-\lambda}e_{m-2}(\lambda)}p.
	\end{equation}

\begin{figure*}[tbh]
\centering
\includegraphics[width=1\textwidth]{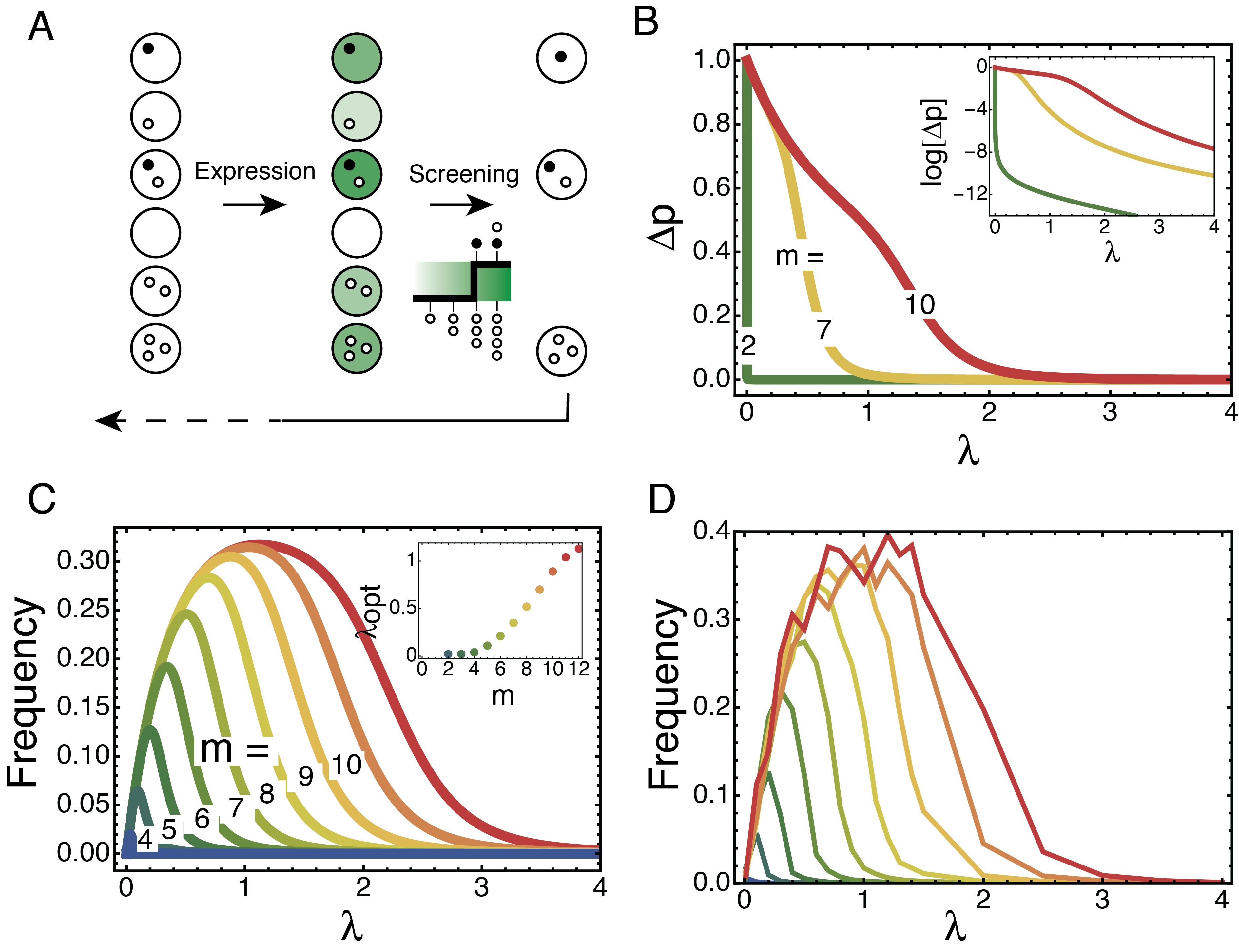}

\caption{Selection dynamics for a step-wise replication function (a.k.a. screening) and two phenotypes, for an initial fraction of improved mutants $p_\mathrm{ini} = 10^{-5}$. {\bf A.}~The process {\bf B.}~Resilience of the selection coefficient to co-infection strongly depends on the activity fold-change $m$. Computed according to (\ref{update-cutoff})  {\bf C.}~Expected frequency of the good variant after one round of screening. When $\lambda$ is small, the low chances of having the improved variant in the population (of size $\lambda M$) dominates the process. The value of $\lambda$ corresponding on average to the highest enrichment ($\lambda_\mathrm{opt}$, inset) increases with the activity increase $m$ brought by the mew mutant. {\bf D.} Numerical simulations for $M = 10^5$, with $p_\mathrm{ini} = 10^{-4}$, each point is averaged over 500 trials.}

\label{fig:3}
\end{figure*}

To look at dynamics, we assume again the apparition of a single mutant in a population of size $N$, and evaluate the frequency change for the first generation. Fig~\ref{fig:3}B shows that the resilience of the step-selection protocols to random co-encapsulation critically depends on $m$: if the improvement brought by the mutant phenotype is only incremental ($m=2$, meaning that the improved mutant is less than 2-fold more active that the wildtype), then the increase in $\lambda$ has a strong impact on the ability to select. On the contrary, if multiple fold changes on activity are expected (e.g. $m = 10$), there exists a range of higher $\lambda$ that bear only mild effects on the dynamics. Here again, one should consider that for a given host budget, larger values of $\lambda$ allow more mutants to be channeled through the selection process. If the process is bottlenecked by the number of available hosts $M$ with a mutation probability $P_\mathrm{ini}$, the probability of observing a favorable mutation is  $P_\mathrm{ini}N = \lambda P_\mathrm{ini}M$. The chances that an improved mutant is actually present in the initial population will control the population fate. Because this probability is small for small $\lambda$, this leads to the emergence of an optimal $\lambda$ that grows with $m$. It can even be larger than one when the activity difference spans a decade (Fig~\ref{fig:3}C). Numerical simulations confirm these predictions and extend their applicability to physically realistic finite populations (Fig~\ref{fig:3}D).

\subsection*{Polymorphic populations}
We have considered so far only simple populations containing just two variants. In most realistic cases however, a larger distribution of activities is available for the variant phenotypes. For example, many RNA viruses exist as quasi-species on a fitness landscape \cite{Domingo:2003cy}; artificial screening processes are applied to libraries of millions of different protein variants, each with a unique activity value \cite{Mair:2017}. We demonstrated in \cite{Zadorin:2017} that the results above can be generalized to the case of a continuous distribution of activities. Moreover, we obtained a general analytical formula to compute the evolution of a population driven by any replication function $f$ for any initial distribution of activities:
	\begin{equation}
	\rho' = \frac{1}{\mathcal N}\left(\sum_{n=0}^\infty \frac{e^{-\lambda}\lambda^n}{(n+1)!}\mathrm{Corr}(\rho^{*n},f)\right)\rho,
	\label{general-rho}
	\end{equation}

\noindent where $\rho$ and $\rho'$ are the probability density function of the activity distribution before and after the selection round, $\mathrm{Corr}$ is the cross-correlation, $\rho^{*n} = \underbrace{\rho\ast\rho\ast\ldots\ast\rho}_{n\text{ times}}$ is the $n$-th convolution power of $\rho$ (we assume here $\rho^{*0} = \delta(x)$, where $\delta(x)$ is the Dirac $\delta$-function, and thus $\mathrm{Corr}(\rho^{*0},f) = f$), and $\mathcal N$ is a normalization coefficient.

In the case where the replication function is linear, the update equation for $\rho$ simplifies to:
	\begin{equation}
	\rho' = \left(1 - g(\lambda) + g(\lambda)\frac{x}{\bar x}\right) \rho,
	\label{update-rho}
	\end{equation}

\noindent where $\bar x$ is the mean of the activity with respect to the distribution $\rho$.

This equation describes a situation where the fitness of an individual depends on the properties of the whole population, and belongs therefore to the class of frequency-dependent selection processes. By frequency-dependent selection we understand the situation when the fitness of a variant depends not only on its own phenotype, but also on its current frequency, on the phenotypes of other variants in the population, and on their frequencies. This becomes clear when the fitness of a phenotype $x$ is computed in the random partitioning context: $\varw_x = g(\lambda) x + \big(1 - g(\lambda)\big)\bar x$.  In the current case of linear replication function, modulation goes through the average activity of the population  $\bar x$. However, frequency dependence is generic for different replication functions at nonzero $\lambda$.

\begin{figure*}[tbh]
\centering
\includegraphics[width=1\textwidth]{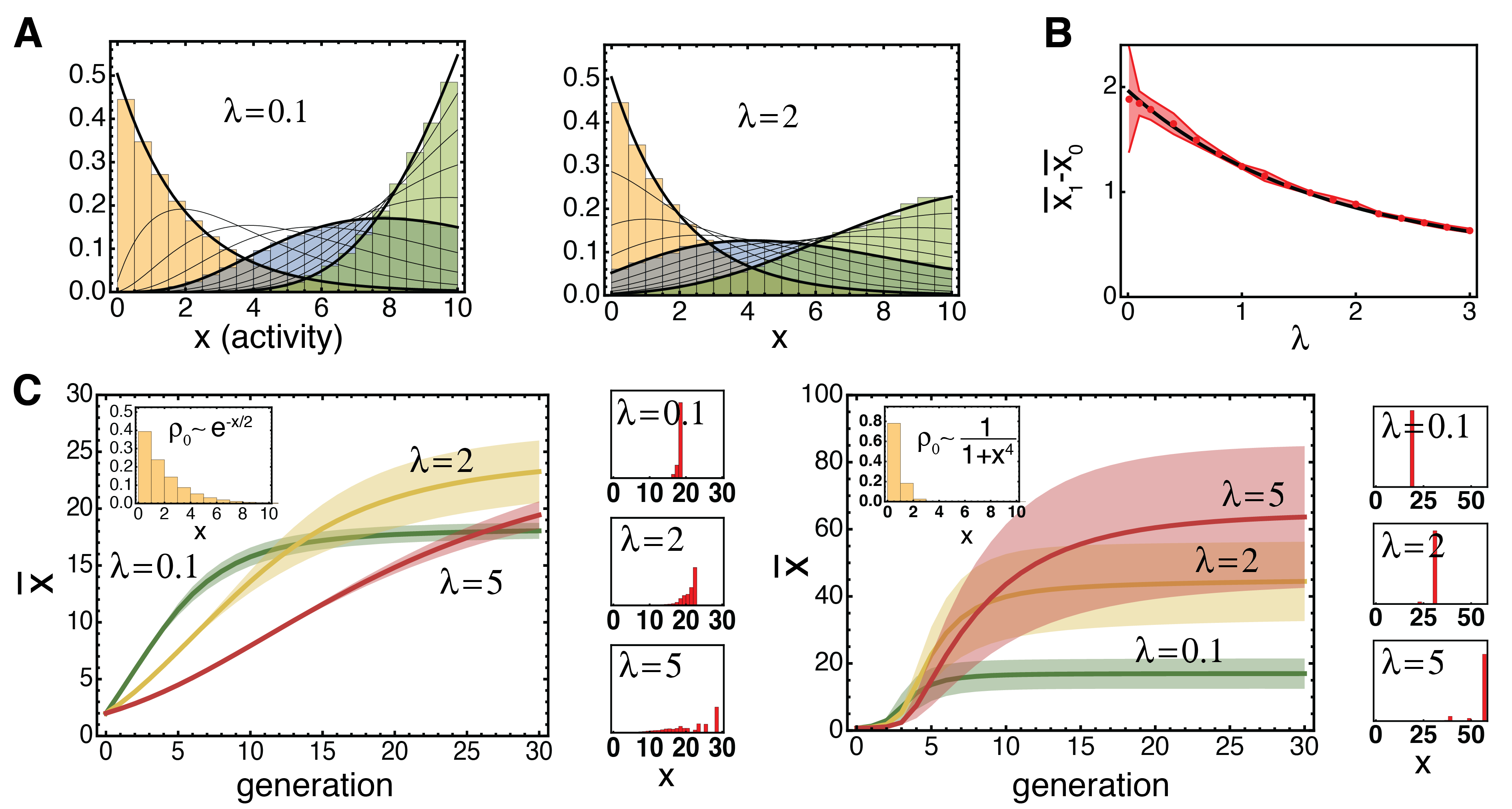}

\caption{Selection on populations with continuous distribution of activity by a linear replication function. {\bf A.}~Dynamic of a population of $10^5$ variants initially selected from an exponential distribution on the interval $[1,10]$. Ten cycles are numerically simulated, with the population  histograms shown for generation 0, 4 and 9. Black lines show the iterated application of (\ref{update-rho}), with thick lines for iterations 0 (exponential distribution), 4 and 9. {\bf B.}~The computed value of the change in mean fitness during the first generation is compared with its prediction $g(\lambda)\cdot\mathrm{const}$, showing a good agreement: the resilience of natural selection to co-infection does not depend on the activity structure of the population. {\bf C.}~Simulated population dynamics starting from populations drawn from unbounded distribution ($10^5$ hosts and various $\lambda$). Mean fitness averaged over 10 runs is shown together with standard deviation (shaded). Insets show the starting distribution. Panels on the right show examples of populations at generation 30.}

\label{fig:4}
\end{figure*}

Equation (\ref{update-rho}) singles out again the mean occupancy  $\lambda$ and the fitness-blurring function $g$ as completely controlling the effect of random multiple encapsulation on the selection process in the linear case. Resilience to random co-encapsulation is therefore a general property, independent of the population polymorphism (Fig~\ref{fig:4}A,B). At constant host resources, and if the initial library distribution has a sufficiently heavy tail, the benefit of a higher mean occupancy emerges naturally: larger populations provide a better sampling of the tail and thus a larger support to the initial activity distribution; since any distribution $\rho$ eventually evolves towards a fixed point $\rho_\infty = \delta(x - x_\text{max})$, where $x_\text{max}$ is the largest value of the support of $\rho$ (so $\rho = 0$ at $x > x_\text{max}$), this advantage eventually (after some generations) overcomes the initial slow-down effect and brings the populations to higher average phenotypic (and thus replicative) values (Fig~\ref{fig:4}C).

The various scenarios observed so far make it clear that, from a replicator population perspective, the best strategy may involve a substantial amount of multiply occupied host compartments. The optimal mean occupancy will primarily depend on properties of the activity distribution of variants, including the frequency of improved variants and the magnitude of the improvement. But, as seen above with screening versus linear selections, it may also be modulated by the shape of the replication function.

\subsection*{Nonlinear replication functions}
\begin{figure}[!tbh]
\centering
\includegraphics[width=.3\textwidth]{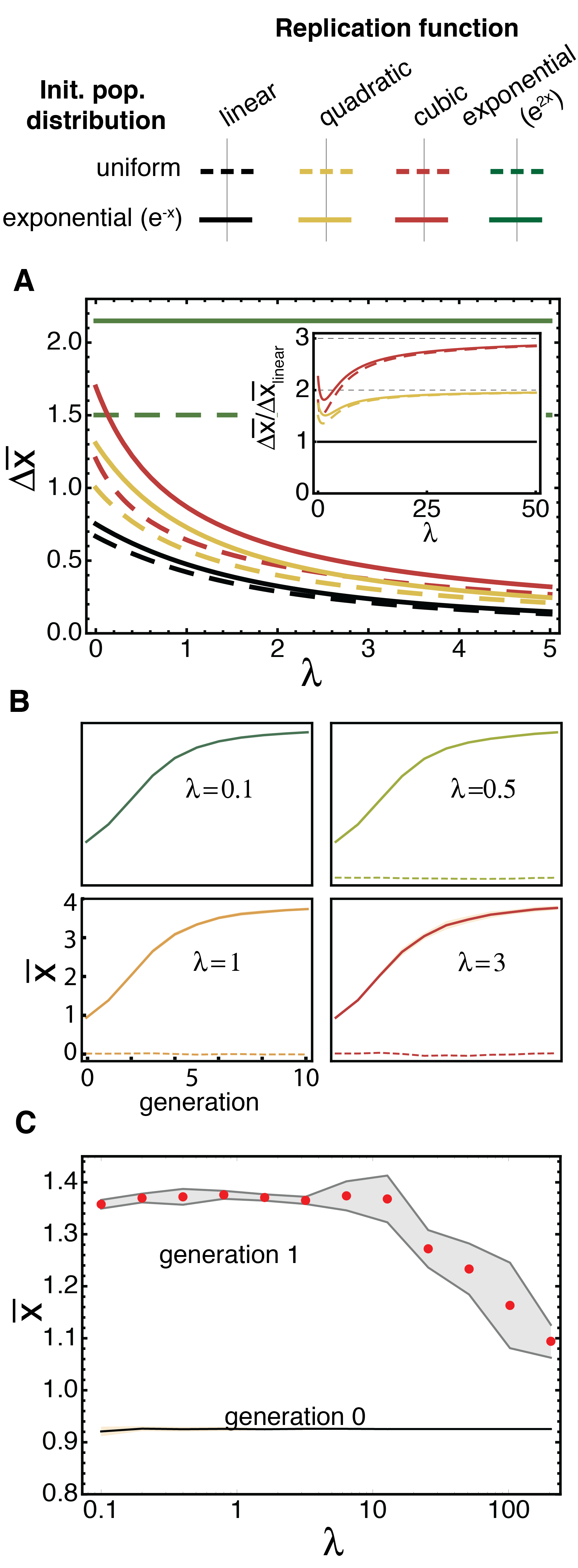}

\caption{{\bf A.}  Mean activity update functions for various activity distributions and replication functions. All distributions are taken on interval $[0,4]$. The inset shows the asymptotic behavior of polynomial selections, relative to the linear case. The exponential replication function is insensitive to random co-compartmentalization, whatever the activity distribution (under the infinite population assumption). Analytical expressions, derived from equation (\ref{general-rho}) are given in Appendix~\ref{appF}. {\bf B.} Numerical simulations starting from the same truncated exponential distribution, using $e^{x/2}$ as a replication function and $10^5$ hosts. For higher  $\lambda$, residuals with  $\lambda=0.1$ are also shown (dashed curve) {\bf C.} Insensitivity to $\lambda$  breaks down at high $\lambda$ when the fraction of total offspring generated by the most loaded host becomes close 1. In this case dynamics gets controlled by stochastic purification effects, which eventually wipe out the natural selection process.}

\label{fig:5}
\end{figure}

We derived the explicit form of equation~(\ref{general-rho}) in the special case of quadratic and cubic replication functions (see Appendix~\ref{appE}). The results of their application, plotted in Fig~\ref{fig:5}A, show that the change in mean fitness of the population after one generation now depends both on mean occupancy and on the shape of the initial distribution. If one plots the polynomial selection efficiency relative to the linear case (Fig~\ref{fig:5}A, inset), a convergence of the ratio to the leading order of the polynomial is observed. In other words, for large $\lambda$, selection by a polynomial function of order $n$ is $n$ times more efficient than selection using linear replication (see Appendix~\ref{appF} for a demonstration).

The observation of gradual improvement with increasingly nonlinear selections prompted us to test the exponential replication function ($f(x) \sim e^{ax}$). In this case (both with and without sharing, see \cite{Zadorin:2017}), equation~(\ref{general-rho}) simplifies to:

	\begin{equation}
	\rho' = \frac{1}{\mathcal N}e^{ax}\rho.
	\end{equation}

The striking implication is that, when the replication process depends exponentially on the phenotypic activity, the selection becomes totally insensitive to  $\lambda$ and random co-compartmentalization (Fig~\ref{fig:5}A). While very counterintuitive, this  effect can be somehow rationalised by considering that the carry over of hitch-hiking mutants with poor phenotypes can now be compensated by the large (multiplicative) fitness benefit that good variants also gain from being co-encapsulated. This result is confirmed by numerical simulations with reasonable population sizes (Fig~\ref{fig:5}B).

However, it is clear that, in physically realistic situations, insensitivity cannot extend to very large $\lambda$. First, the finite resources available in each host put a higher bound on the number of offspring that it can support. Second, a more subtle effect develops because the distribution of total activity in compartments spreads more and more with increasing $\lambda$; at some point, the offspring generated in the most loaded host, due to the exponential replication law, will start to actually represent a significant fraction of the total new population. This effect will prematurely deplete the population from its diversity and therefore negatively affect the selection of the best mutants. In practice, there is therefore a threshold above which multiple occupancy starts again to affect selection, this time not because of fitness blurring, but because of stochastic purification effects inherent to the exponential growth (Fig~\ref{fig:5}C).

\begin{figure}[!tbh]
\centering
\includegraphics[width=.3\textwidth]{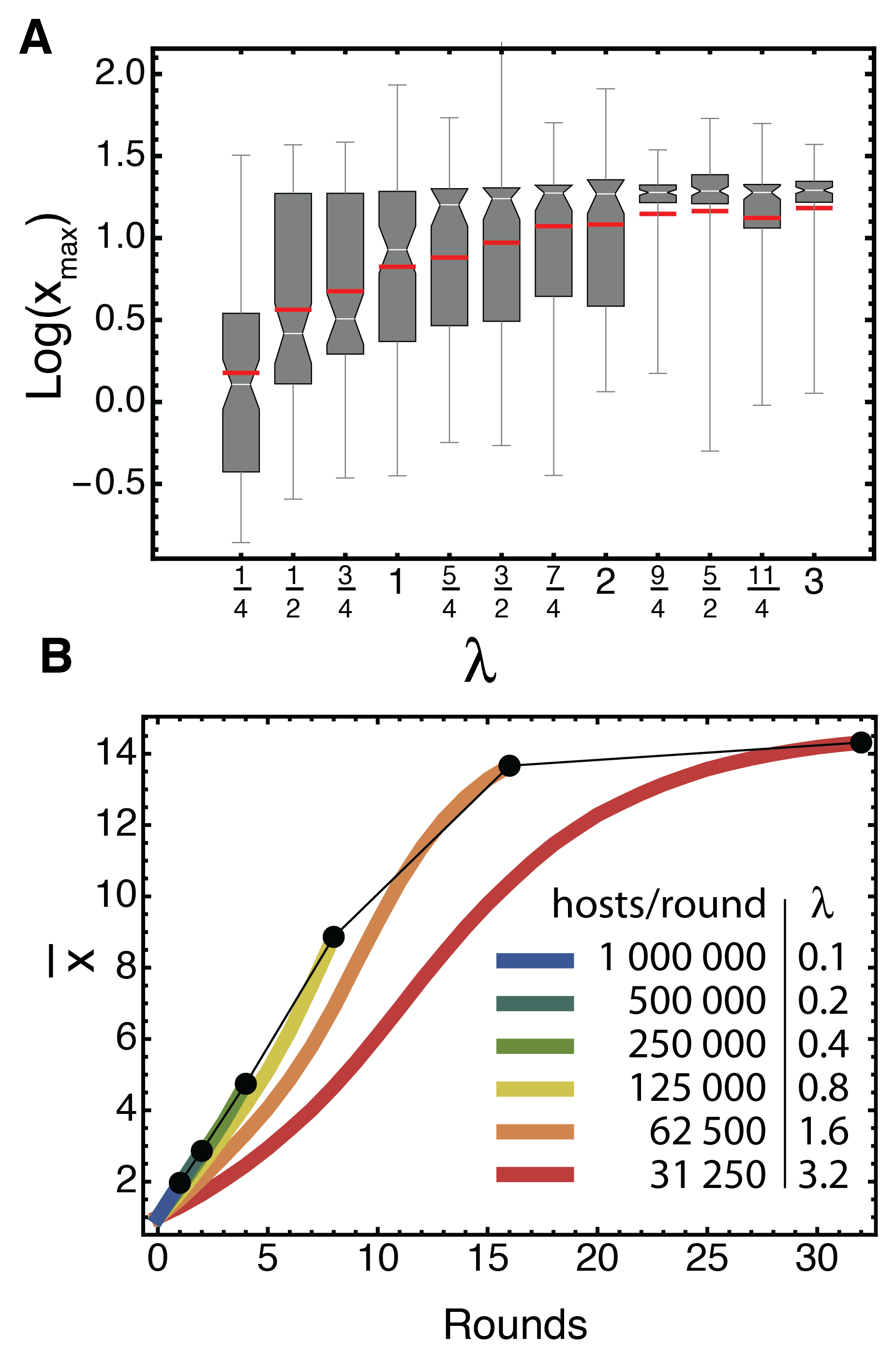}

\caption{{\bf A.} Evolution on a $NK$-landscape at various $\lambda$. ($N=20$, $K=3$, 20 letters alphabet; see Appendix~\ref{appH}). For selection, the total activity is shared and the replication function is linear. Up to four random mutations are applied to each individual at each round. The whisker chart gives the best variant found in the population at round 10 (100 runs with 1000 hosts are performed for each $\lambda$, each stochastic run initiated from the same, non-lethal seed individual, and on the same landscape). The red bar is the mean. {\bf B.} Selections with limited resources. A population of $10^5$ random variants, drawn from an exponential distribution, are selected using a total budget of $10^6$ hosts. Change in mean activity of the population for different protocols using 1 to 32 rounds (thus  $10^6$ to 32250 hosts per round) is shown. Linear replication and activity sharing are assumed here also.}

\label{fig:6}
\end{figure}

Our discussion is based on the case where all variants are present in the initial population, and no new diversity is introduced during the selection process. In an evolutionary context, our results will hold if the mutation rate is small compared to the selection rate (for example, in the case of linear selection from two phenotypes, if mutation probability per generation is smaller than the inverse characteristic time of the mutation sweep, $t_\mathrm{tot} = t_1 + t_2$ defined by equations~(\ref{t1})) \cite{Bershtein:2008}. More generally, one may still anticipate that high $\lambda$ strategies, that are advantageous for selections, may also be beneficial in an evolutionary walk. A formal treatment of this question would require informed hypotheses on the statistical structure of the underlying landscape \cite{Firnberg:2014iq,Domingo:2003cy,Bershtein:2006rob,Hietpas:2011exp}. However, we can still explore numerically this possibility on model landscapes. NK landscapes provide a model of fitness landscape where the degree of epistasis can be adjusted via a single parameter K, and N is the length of the genome. Increasing K makes these landscapes gradually more rugged and therefore harder to navigate via evolutionary strategies. Fig~\ref{fig:6}A shows simulated evolution runs with random compartmentalisation on a modified (sparse) $N\!K$-model \cite{Kauffman:1987tow,Kauffman:1989ada,Kauffman:1989nk} (see Appendix~\ref{appH} for details). In this case, we observe that runs using higher $\lambda$ are indeed able to discover on average better mutants, most probably by leveraging their higher throughput.

\section*{Discussion}
We have found the slowdown effect associated with multiple occupancy, already noted by others \cite{Baret:2009,Novella:2004ec}, to be less deteriorating than intuitively anticipated. Additionally, it critically depends on the shape of the replication function and may completely vanish in the important case of exponential replicators \cite{Rosinski:2002,Lu:2011,Lui:2013,Ghadessy:2001hj} with additive Malthusian fitness, where higher $\lambda$ appear to have no cost. Our analytical results were derived under the assumption of the Poisson partitioning of infinite populations, but they appear to be robust to stochastic effects in smaller, realistic populations. Tolerating higher $\lambda$-values means that larger populations of replicators can be channeled through selection processes. In a world of limited host ressources and immense sequence spaces, this increases the chance of finding rare beneficial mutations. These findings may therefore have strong implications for both natural and artificial evolution. More generally, we note that similar stochastic co-selection effects occur also in a variety of related cases. It can happen for example because of non-specific aggregation of the individual variants in the absence of compartments; di- or polyploidy of many organisms, as well as spatial population structure in the presence of a common good also lead to frequency dependent outcomes. These situations could be covered using simple modifications of our model \cite{Gerstein:2009,Cressman:2013}. 

Artificial directed evolution protocols select polymers with desirable properties out of large librairies of variants. In order to optimise the selection stringency, many high-throughput, \emph{in vitro} experimental designs try to maintain at most single phenotype per host by working at low $\lambda$ \cite{Griffiths:2003dir}. This in turn imposes a limit on the size of the library that can be evaluated.  Our results suggest that using libraries size exceeding the number of hosts could in many cases be more favorable than prematurely bottlenecking the population diversity. First, the slowdown effect on the selection of best variants is not as bad as one could expect. With an $n/\lambda$ asymptotic for polynomial selection laws, increasing the occupancy will never break the ability to select more than it increases the throughput. Additionally, it will be alleviated by the iterated nature of most selection processes. For example Fig~\ref{fig:6}B shows that, given available experimental resources expressed as a total budget of hosts (over all rounds), and a library size, the best strategy to obtain the highest final mean activity for the selected population is always to split the host budget in as many rounds as possible (see Appendix~\ref{appG}). In an experiment constrained to a limited number of rounds, knowledge of the population phenotypic structure and the shape of the replication function should help one define the best $\lambda$ schedule.

From a biological perspective, infectious viral occupancy has often been considered from the host cells' standpoint: only when the ratio of virus to cell is relatively high ($> 3$)  can one consider that most cells are indeed infected \cite{Kutter:2004}. It has much less often been looked at from the virus' standpoint \cite{Frank:2001}, even if it is know that multiplicities of infection as high as 10--100 can be observed. Novella \emph{et al.} considered a related simple case with two phenotypes and suggested that the phenotype-averaging effect of higher occupancies can benefit the population by sustaining diversity in the face of changing selection pressures \cite{Novella:2004ec}. Here, without considering changing environments, we show that favouring multiple occupancy can actually be a valid strategy to optimise coverage of sequence space and select fitter mutants. High occupancy strategies become especially interesting in two cases: i) when the distribution of fitness effect is skewed enough: in a rich neighbourhood of the fitness landscape, selection with high replicator occupancy will ultimately lead to better improvements than strategies attempting to limit it; ii) when the replication mechanism leads to nonlinear or exponential growth dynamics, a case which is common for viral replicators \cite{Rosinski:2002,Lu:2011,Lui:2013}. As some viruses and plasmids have specific mechanisms to control their replication parameters \cite{Ptashne:2004}, this opens the intriguing perspective that the control of the mean occupancy could itself be the subject of evolution.

Finally, the effect of stochastic purification, observed for exponential selection, deserves a separate attention. Unlike the classical genetic drift, this effect gets \emph{stronger} when the population size \emph{increases}. Furthermore, Figure~\ref{fig:5}C suggests a behaviour that resembles a phase transition rather than a smooth change. All this indicates that we deal with a new phenomenon in evolutionary dynamics.

\appendix


\section{Derivation of equation (\ref{g-linear-update}): linear replication function with sharing of the replication activity}
\label{appA}
\subsection{Discrete time dynamics}

We will first treat the case of a linear replication function with sharing of the replication activity in a compartment. This corresponds to the 
replication function $f(x) = a x$, where $a$ is a constant that defines the overall replication efficiency. Without loss of generality, we will 
assume $a = 1$ and thus $f(x) = x$.

We remind that under random partitioning assumption with the mean occupancy $\lambda$, the probability to find an $ij$-compartment (see the main text) 
is assumed to be
	\begin{equation}
	P_{ij} = \frac{e_{}^{-\lambda}\lambda^{i+j}}{(i+j)!} C^i_{i+j} p^i (1-p)^j = \frac{e_{}^{-\mu}\mu^i}{i!}\frac{e_{}^{-\nu}\nu^j}{j!},
	\end{equation}

\noindent where $C^k_n = n!/(k! (k-n)!)$ is the binomial coefficient and, as in the main text, $\mu \deq p\lambda$ and $\nu \deq (1-p)\lambda$, so $\nu + \mu = \lambda$.

With sharing of the replication activity, each initial replicator (of phenotype 1 or of phenotype 2 alike) in an 
$ij$-compartment has $\varw_{ij} = (ix_1 + jx_2)/(i+j)$ copies passed to the next generation (this is the local fitness of the compartment). The fitness of 
phenotype 1 in the whole population is then equal to

\begin{equation}
\varw_1 = \frac{\sum\limits^{\infty}_{i=1}\sum\limits^{\infty}_{j=0} P_{ij}i \varw_{ij}}
{\sum\limits^{\infty}_{i=1}\sum\limits^{\infty}_{j=0} P_{ij}i} =
\frac{1}{\mu}\sum_{i=1}^\infty\sum_{j=0}^\infty e^{-\lambda}\frac{\mu^i \nu^j}{i!j!} i \frac{ix_1 + jx_2}{i+j} =
\sum_{i=1}^\infty\sum_{j=0}^\infty e^{-\lambda}\frac{\mu^{i-1} \nu^j}{(i-1)!j!} \frac{ix_1 + jx_2}{i+j} =
\sum^{\infty}_{i=0}\sum^{\infty}_{j=0}P_{ij}\frac{ix_1 + jx_2 + x_1}{i + j + 1},
\end{equation}

\noindent where we used the identity $\sum\limits^{\infty}_{i=1}\sum\limits^{\infty}_{j=0}P_{ij}i = p\lambda = \mu$ and the change of index $i \mapsto 
i - 1$. In the same way we obtain

\begin{equation}
\varw_2 = \sum^{\infty}_{i=0}\sum^{\infty}_{j=0}P_{ij}\frac{ix_1 + jx_2 + x_2}{i + j + 1},
\end{equation}

\noindent and therefore the mean fitness of the population is equal to

\begin{multline}
\bar \varw = p\varw_1 + (1-p)\varw_2 =\\
=\frac{e_{}^{-\lambda}}{\lambda}\left(\sum^{\infty}_{i=1}\sum^\infty_{j=1}\frac{\mu^i\nu^j}{i!j!}(ix_1+jx_2) +
x_1\sum^\infty_{i=1}\frac{\mu^i}{i!}i + x_2\sum^\infty_{j=1}\frac{\nu^j}{j!}j\right) =\\
=\frac{e_{}^{-\lambda}}{\lambda}\left(x_1\mu e_{}^\mu(e_{}^\nu-1) + x_2\nu e_{}^\nu(e_{}^\mu-1) + x_1\mu e_{}^\mu + x_2\nu e_{}^\nu\right) =
px_1 + (1-p)x_2 = \bar x,
\end{multline}

\noindent where $\bar x$ is the mean activity in the initial population. 
Note that the mean fitness is exactly equal to the mean activity. This property is a specific feature of the linear selection function with sharing. Indeed, under this condition, the local fitness for any replicator in any compartment is exactly equal to the mean activity in that compartment.

The last expression to be computed is

\begin{multline}
\varw_1 - \varw_2 = (x_1 - 
x_2)\sum^\infty_{i=0}\sum^\infty_{j=0}e_{}^{-\lambda}\lambda^{i+j}\frac{p^i(1-p)^j}{i!j!}\frac{1}{i+j+1}=\\
=(x_1 - x_2)\sum^\infty_{n=0}e_{}^{-\lambda}\frac{\lambda^n}{n!}\frac{1}{n+1} = (x_1 - x_2)\frac{1 - e_{}^{-\lambda}}{\lambda}.
\end{multline}

As in the main text, we will give a separate notation to the factor that depends on $\lambda$

\begin{equation}
g(\lambda) \deq \frac{1 - e_{}^{-\lambda}}{\lambda}.
\end{equation}

Finally, the update equation for the frequency $p$ takes the form

\begin{equation}
\Delta p = g(\lambda)\frac{p(1-p)(x_1 - x_2)}{px_1 + (1-p)x_2}.
\label{g-linear-update-app}
\end{equation}

It is worth to note that the expression for $\Delta p$ is the same as in the classical population genetics, under assumption that the activity defines 
fitness directly, except for an additional factor $g(\lambda)$, that effectively slows down the selection process. This can be easily seen from taking 
the limit $\lambda \to 0$, that corresponds to a single individual in each occupied compartment. But this also can be directly demonstrated. Indeed, in 
the classical case, the update equation is given by
 
\begin{equation}
p' = p\frac{x_1}{\bar x} = \frac{px_1}{px_1 + (1-p)x_2} \quad \text{or} \quad \Delta p = \frac{p(1-p)(x_1 - x_2)}{px_1 + (1-p)x_2}.
\end{equation}

Equation (\ref{g-linear-update-app}) is not solvable in closed form. Nevertheless, it is still possible to obtain some qualitative and quantitative 
estimates without solving the dynamics.

If $p$ is not yet fixed ($p \neq 0$, $p \neq 1$), it grows (i.e., $\Delta p > 0$) if and only if $x_1 > x_2$, and it decays (i.e. $\Delta p < 0$) if 
and only if $x_1 < x_2$. This is obvious from (\ref{g-linear-update-app}) provided $0 < p < 1$. $p = 0$ and $p = 1$ are fixed points of the dynamics. 1 is 
attracting and 0 is repelling when $x_1 > x_2$ and they change their stability when $x_1 < x_2$. So, the mutant with the best activity is always 
selected. However, the increase of $\lambda$ decreases the rate of this selection. For instance, for high $\lambda$, the change in $p$ at every step is 
approximately $1/\lambda$ of what it would be in the classical population genetics case.

In the following, we will assume $x_1 > x_2$. When $p$ is very small, we can approximate (\ref{g-linear-update-app}) with its linearization

\begin{equation}
p' = \left(1 + g(\lambda)\frac{x_1 - x_2}{x_2}\right)p = \alpha p.
\label{g-linearized-app}
\end{equation}

\noindent Then, for small $p$, the solution is approximated by $p_t = \alpha^t p_0$, where $p_t$ is the value of $p$ at time $t$, where $t$ is 
measured in number of generations. On the other hand, when $p \approx 1$, (\ref{g-linear-update-app}) can be substituted with its linearization about $1 - 
p$

\begin{equation}
1 - p' = \left(1 - g(\lambda)\frac{x_1 - x_2}{x_1}\right)(1 - p) = \beta (1-p).
\label{g-linearized2-app}
\end{equation}

\noindent The solution near 1 can be then approximated by $1 - p_t = \beta^t (1 - p_0)$.

There is a particular case when (\ref{g-linearized-app}) is not applicable, that is $x_2 = 0$ (phenotype 2 is ``lethal''). In this case linear 
approximation (\ref{g-linearized2-app}) becomes exact, as it is easily seen from (\ref{g-linear-update-app}). Interestingly, if $\lambda = 0$, phenotype 1 is 
fixed in one generation, so the dynamics is not reversible. This is manifested by $\beta = 0$. Non-zero value of $\lambda$ makes this process stretched 
in time, as $\beta > 0$.

The values of $\alpha$ and $\beta$ can be used to approximate the whole dynamics by a piecewise-linear one. For example, if the real population size is 
$N$, then the condition $\alpha^{t_1} p_0 = 1$, where $p_0 = 1/N$ provides an estimation on the time for a population with a single individual of 
phenotype 1 to reach a stage where this phenotype has a macroscopic representation, i.e. the number of its individuals is comparable with $N$ (we 
completely ignore the genetic drift here). This time can be called the invasion time (of phenotype 1 into a population of phenotype 2), and it is given 
by

\begin{equation}
t_1 = \frac{\ln N}{\ln \alpha}.
\end{equation}

In the same way, the condition $\beta^{-t_2}q_f = 1$, where $q_f = 1 - p_f = 1/N$ gives an estimate of the time it takes for a macroscopically present 
phenotype 1 to be finally fixed, which is approximated here by the time to reach the frequency $p_f = (N-1)/N$ (only one individual of phenotype 2 is left). 
This results in

\begin{equation}
t_2 = -\frac{\ln N}{\ln \beta}.
\end{equation}

The time $t_\mathrm{tot} = t_1 + t_2$ gives a typical total time for the phenotype 1, superior to the phenotype 2, to be fixed starting from one individual 
(completely ignoring any stochasticity). This is the time of a selection sweep by a single mutation.

Below, we present an alternative, simpler way to derive equation~(\ref{g-linear-update}) which however relies on some advanced theory. As shown in our 
accompanying work \cite{Zadorin:2017}, if the fitness per individual in a compartment with $n$ individuals of activities $x_1$, \ldots, $x_n$ (not 
necessarily different) is given by the formula $\varw(x_1,\ldots,x_n) = \phi(n)\sum_{i=1}^n x_i$, where $\phi$ is some function of $n$ that represents some 
specific features of the selection process like sharing, then the population-wide fitness of a phenotype $k$, $\varw_k$, can be found as

\begin{equation}
\varw_k = \frac{1}{\lambda}\sum_{n=1}^\infty \frac{e^{-\lambda} \lambda^n}{n!} n \phi(n) \big(x_k + (n-1)\bar x\big) =
\sum_{n=0}^\infty \frac{e^{-\lambda} \lambda^n}{n!}\phi(n+1)(x_k + n \bar x),
\label{fk-general-app}
\end{equation}

\noindent where $\bar x$ is the population average activity. In the case considered here, $\bar x = p x_1 + (1-p) x_2$. The population average fitness 
can be found as

\begin{equation}
\bar \varw = \frac{\bar x}{\lambda}\sum_{n=1}^\infty \frac{e^{-\lambda} \lambda^n}{n!} n^2 \phi(n) =
\bar x \sum_{n=0}^\infty \frac{e^{-\lambda} \lambda^n}{n!} (n+1)\phi(n+1).
\label{f-general-app}
\end{equation}

In our case, $\phi(n) = 1/n$ and, as $x_k + n\bar x = (n+1)\bar x + x_k - \bar x$ and taking into account that $\sum_{n=0}^\infty 
e^{-\lambda}\lambda^n/n! = 1$ and $\sum_{n=0}^\infty e^{-\lambda}\lambda^n/(n+1)! = g(\lambda)$, we can easily conclude that

\begin{equation}
\varw_k = \bar x + (x_k - \bar x) \sum_{n=0}^\infty \frac{e^{-\lambda} \lambda^n}{n!}\frac{1}{n+1} = \bar x + g(\lambda)(x_k - \bar x)
\quad \text{and} \quad
\bar \varw = \bar x,
\end{equation}

\noindent from which all the rest follows. In the following, we will use this simple way to derive update equations, when possible, to avoid cumbersome 
computations.

\subsection{Continuous time dynamics}
\label{cont-time}

If the change in $p$ is small at every generation in comparison with the actual value of $p$, (\ref{g-linear-update}) can be substituted by a 
differential equation, which is much easier to deal with. If the time $t$ is measured in generations, the corresponding differential equation reads

\begin{equation}
\frac{dp}{dt} = g(\lambda)\frac{p(1-p)(x_1 - x_2)}{px_1 + (1-p)x_2}.
\end{equation}

This equation is solvable in closed implicit form and its solution for $p(0) = p_0$ is given by

\begin{equation}
\left(\frac{p}{p_0}\right)^{x_2}
\left(\frac{1-p_0}{1-p}\right)^{x_1} =
e_{}^{g(\lambda)(x_1 - x_2)t}.
\end{equation}

This solution describes a sigmoidal curve with different asymptotic behavior at different sides. When $t \to -\infty$, $p \sim e_{}^{at}$, where $a = 
g(\lambda)(x_1-x_2)/x_2$. When $t \to +\infty$, $1 - p \sim e_{}^{bt}$, where $b = g(\lambda)(x_2 - x_1)/x_1$. As $(1 + x)^t = e_{}^{xt} + o(x)$ and 
$\alpha = 1 + a$, $\beta = 1 + b$, This agrees well with the asymptotic behavior of the discrete time case obtained above. Here, the effect of the 
random partitioning is even more evident: the selection slowdown is directly expressed by the multiplication of the characteristic rates by 
$g(\lambda)$. The estimates for the phenotype 1 invasion and the fixation times are now given by

\begin{eqnarray}
t_1 = \dfrac{\ln N}{a} =& \dfrac{x_2 \ln N}{g(\lambda)(x_1 - x_2)},\\
t_2 = -\dfrac{\ln N}{b} =& \dfrac{x_1 \ln N}{g(\lambda)(x_1 - x_2)}.
\end{eqnarray}

\section{Derivation of the update equation for the linear case without sharing and for the case of sharing both the activity and offspring}
\label{appB}
When there is no sharing of the offspring and every replicator in a compartment acquires the number of copies that is equal to the total activity in 
the compartment given by the replication function $f(x) = x$. The only difference with the previous case is that the local fitness in an 
$ij$-compartment is now given simply by $\varw_{ij} = ix_1 + jx_2$ instead of $(ix_1 + jx_2)/(i+j)$. In terms of equations~(\ref{fk-general-app}) 
and~(\ref{f-general-app}), we have $\phi(n) = 1$. Thus, we find that

\begin{equation}
\varw_k = x_k + \bar x \sum_{n=0}^\infty \frac{e^{-\lambda} \lambda^n}{n!}n = x_k + \lambda \bar x
\end{equation}

\noindent and

\begin{equation}
\bar \varw = \bar x \sum_{n=0}^\infty \frac{e^{-\lambda} \lambda}{n!}(n + 1) = (\lambda + 1)\bar x.
\end{equation}

Therefore, the update equation takes the form

\begin{equation}
\Delta p = p(1-p)\frac{x_1 - x_2}{(\lambda + 1)\bar x} = \frac{1}{1 + \lambda}\,\frac{p(1-p)(x_1 - x_2)}{p x_1 + (1-p) x_2}.
\end{equation}

This result differs from equation~(\ref{g-linear-update}) only by the slowdown factor. Therefore, all the results from the previous section are 
literally correct for the selection without sharing, if $g(\lambda)$ is replaced everywhere by the function $h(\lambda) \deq 1/(1 + \lambda)$.

Let us now show how to use this method to obtain the results of Novella \emph{et al.} from \cite{Novella:2004ec}. Their analysis correspond to the case when not only the total number of offspring is shared but the total activity in a compartment is also defined as the average activity of individuals in the compartment. This case 
corresponds to $\phi(n) = 1/n^2$. Indeed, the average activity is given by $X =( \sum_{i=1}^n x_i)/n$ and every individual of the compartment obtains 
$X/n$ copies. Therefore, we can immediately conclude that

\begin{equation}
\varw_k = (x_k - \bar x) \sum_{n=0}^\infty \frac{e^{-\lambda} \lambda^n}{n!} \frac{1}{(n+1)^2} + 
\bar x \sum_{n=0}^\infty \frac{e^{-\lambda} \lambda^n}{n!} \frac{1}{n + 1} =
\frac{\Ei(\lambda) - \ln \lambda -\gamma}{\lambda e^\lambda} (x_k - \bar x) + g(\lambda) \bar x,
\end{equation}

\noindent where $\Ei$ is the so called exponential integral

\begin{equation}
\Ei(x) = -\inth{-x}{\infty}\frac{e^{-y}}{y}\,dy
\end{equation}

\noindent and $\gamma = 0.5772\ldots$ is the Euler constant. This formula correspond exactly to the one derived in 
\cite{Novella:2004ec}.

Likewise,

\begin{equation}
\bar \varw = g(\lambda)\bar x.
\end{equation}

Therefore, the update equation is given by

\begin{equation}
\Delta p = \frac{\Ei(\lambda) - \ln \lambda - \gamma}{e^\lambda - 1}\,\frac{p(1-p)(x_1 - x_2)}{px_1 + (1-p)x_2},
\end{equation}

\noindent and thus, all the conclusions of the previous section literally apply to the case of activity averaging with offspring sharing, if the 
function $g(\lambda)$ is everywhere replaced with the function $\psi(\lambda) = \big(\Ei(\lambda) - \ln\lambda - \gamma\big)/(e^\lambda - 1)$.

\section{Derivation of equation (\ref{update-cutoff}): cut-off case without sharing}
\label{appC}

Let us consider the case of the cut-off replication function, which corresponds to the artificial screening process applied to the selection of, for 
example, improved enzyme variants. Sharing does not happen in screening. The selection function $f$ has the form of a step function $f(x) = 
\theta(x-x_1)$, where $\theta$ is the Heaviside function. Here we assume $\theta(0) = 1$, so the function $\theta$ is left-continuous. This means that 
all individual of a compartment are passed to the next round, if the activity in the compartment is not lower than $x_1$. This, in turn, means that the 
selection threshold is set exactly at the activity $x_1$. An important parameter is the number $m = \lceil x_1/x_2 \rceil$ such that $(m-1) x_2 < x_1$ 
and $m x_2 \geqslant x_1$. The meaning of $m$ is the minimal number of phenotype~2 individuals that are required in one compartment to trigger 
selection in the absence of any phenotype~1. Under such conditions, the local fitness in an $ij$-compartment is equal to

\begin{equation}
\varw_{ij} = \begin{dcases}
1,& i > 0\; \text{or}\; j \geqslant m,\\
0,& i = 0\; \text{and}\; j < m.
\end{dcases}
\end{equation}

The population-wide fitness of phenotype 1 is given by

\begin{equation}
\varw_1 = \frac{1}{\mu}\sum_{i=1}^{\infty}\sum_{j=0}^{\infty}i P_{ij} = 1,
\end{equation}

\noindent which is intuitively clear, while the fitness of phenotype 2 is equal to

\begin{equation}
\varw_2 = \frac{1}{\nu}\sum_{i=1}^{\infty}\sum_{j=1}^{\infty}j P_{ij} + 
\frac{1}{\nu}\sum_{j=m}^{\infty}\frac{je^{-\lambda}\nu^j}{j!} =
1 - \frac{e^{-\lambda}}{\nu}\sum_{j=1}^{m-1}\frac{j\nu^j}{j!} =  1 - e^{-\lambda}\sum_{j=0}^{m-2}\frac{\nu^j}{j!}= 1 - e^{-\lambda}e_{m-2}(\nu),
\end{equation}

\noindent where $e_n$ is the truncation of the Taylor series of the exponential function to the $n$-th term, so

\begin{equation}
e_n(x) \deq \sum_{k=0}^n \frac{x^k}{k!}.
\end{equation}

The difference between these values is equal to

\begin{equation}
\varw_1 - \varw_2 = e^{-\lambda}e_{m-2}(\nu),
\end{equation}

\noindent and the population average fitness is equal to

\begin{equation}
\bar \varw =
p\varw_1 + (1-p)\varw_2 = p + (1-p)(1 - e^{-\lambda}e_{m-2}(\nu)) = 1 - (1-p)e^{-\lambda}e_{m-2}(\nu).
\end{equation}

This allows to write the difference equation for the selection dynamics

\begin{equation}
\Delta p = p(1-p)\frac{e_{m-2}\big((1-p)\lambda\big)}{e^\lambda - (1-p)e_{m-2}\big((1-p)\lambda\big)}.
\label{m-update-app}
\end{equation}

As one can easily see, for any $p$ such that $0 < p < 1$, $\Delta p > 0$, whence $0$ and $1$ are stationary points, as it was the case with the 
linear selection. The limit case of a lethal phenotype 2 corresponds to $m \to \infty$ and is described by the equation

\begin{equation}
\Delta p = p(1-p)\frac{e^\nu}{e^\lambda - (1-p)e^\nu}.
\end{equation}

The asymptotic behavior again can be analyzed by the linearization of the dynamic equation about $p = 0$ and $p = 1$. At the limit $p \to 0$ the 
linearized update equation is

\begin{equation}
p' = \frac{1}{1 - e^{-\lambda}e_{m-2}(\lambda)}p = \alpha p,
\end{equation}

\noindent while at $p \to 1$ the linearization is

\begin{equation}
1 - p' = (1 - e^{-\lambda})(1-p) = \beta (1-p),
\end{equation}

\noindent and it does not depend on $m$.

The values of $\alpha$ and $\beta$ can be used for estimations of the characteristic times. Note that, at $\lambda = 0$, $\beta$ becomes 0. In fact, at 
this value of $\lambda$ the dynamics is non-reversible, as seen from (\ref{m-update-app}). Indeed, this equation becomes $\Delta p = 1 - p$ meaning that 
the fixation of phenotype 1 happens in a single step. This is not surprising, as this case corresponds to a classical selection from two phenotypes, 
one of which is lethal ($\lambda \to 0$ corresponds to an individual selection). As in the case of a linear replication function, the limit of a lethal 
phenotype~2 ($m \to \infty$) corresponds to $\alpha \to \infty$, so the invasion of a viable phenotype into a population of lethal phenotypes happens 
effortlessly.

If the screening threshold value $x_0$ is lower than $x_1$ (but, of course, higher than $x_2$), and the replication function is $f(x) = \theta(x - 
x_0)$, the formulas stay the same except that now one has to use $m = \lceil x_0/x_2 \rceil$ instead of $\lceil x_1/x_2 \rceil$.

\section{Derivation of the update equation for the cut-off case with sharing}
\label{appD}

Let us consider the previous case, but with sharing. The replication function $f$ has a form of a step function $f(x) = a\theta(x-x_1)$, where $\theta$ 
is the Heaviside function. We will assume $\theta(0) = 1$, so the function $\theta$ is left-continuous. We will use the same parameter $m = \lceil 
x_1/x_2 \rceil$ as in Appendix~\ref{appC} such that $(m-1)x_2 < x_1$ and $m x_2 \geqslant x_1$. Under such conditions, the local fitness 
in an $ij$-compartment is equal to

\begin{equation}
\varw_{ij} = \begin{dcases}
\frac{a}{i+j},& i > 0\; \text{or}\; j \geqslant m,\\
0,& i = 0\; \text{and}\; j < m.
\end{dcases}
\end{equation}

\noindent Here $a$ is some large number (total number of copies produced in a compartment with functional mix) and we assume that $a/(i+j)$ makes 
sens as an average number of copies over all compartments of $ij$-composition.

Then the mean population fitness of phenotype 1 is given by

\begin{equation}
\varw_1 = \frac{a}{\mu}\sum_{i=1}^{\infty}\sum_{j=0}^{\infty}P_{ij}\frac{i}{i+j} = a\sum_{i=0}^{\infty}\sum_{j=0}^{\infty}
\frac{P_{ij}}{i + j + 1} = a g(\lambda),
\end{equation}

\noindent while the mean population fitness of phenotype 2 is

\begin{equation}
\varw_2 = \frac{a}{\nu}\sum_{i=1}^{\infty}\sum_{j=1}^{\infty}P_{ij}\frac{j}{i+j} + 
\frac{a}{\nu}\sum_{j=m}^{\infty}\frac{e^{-\lambda}\nu^j}{j!} =
\varw_1 - \frac{a e^{-\lambda}}{\nu}\sum_{j=1}^{m-1}\frac{\nu^j}{j!} = \varw_1 - \frac{a e^{-\lambda}}{\nu}(e_{m-1}(\nu) - 1),
\end{equation}

\noindent where $e_n$ is again the truncation of the Taylor series of the exponential function to the $n$-th term and $\nu = (1-p)\lambda$. The 
difference between these values is

\begin{equation}
\varw_1 - \varw_2 = \frac{a e^{-\lambda}}{\nu}(e_{m-1}(\nu) - 1),
\end{equation}

\noindent and the mean population fitness is equal to

\begin{equation}
\bar \varw =
p\varw_1 + (1-p)\varw_2 = \frac{a e^{-\lambda}}{\lambda}(e^\lambda - e_{m-1}(\nu)).
\end{equation}

This allows to write the difference equation for the selection dynamics

\begin{equation}
\Delta p = p\frac{e_{m-1}(\nu) - 1}{e^\lambda - e_{m-1}(\nu)} = p\frac{e_{m-1}((1-p)\lambda) - 1}{e^\lambda - e_{m-1}((1-p)\lambda)}.
\label{m-update2-app}
\end{equation}

As one can easily see, for any $p$ such that $0 < p < 1$, $\Delta p > 0$, whence $0$ and $1$ are stationary points, as it was the case with the 
linear selection. The limit case of a lethal phenotype 2 corresponds to $m \to \infty$ and is described by the equation

\begin{equation}
\Delta p = p\frac{e^\nu - 1}{e^\lambda - e^\nu}.
\end{equation}

The asymptotic behavior again can be analyzed by the linearization of the dynamic equation about $p = 0$ and $p = 1$. At the limit $p \to 0$ the 
linearized update equation is

\begin{equation}
p' = \frac{e^\lambda - 1}{e^\lambda - e_{m-1}(\lambda)} p = \alpha p,
\end{equation}

\noindent while at $p \to 1$ the linearization is

\begin{equation}
1 - p' = \left(1 - \frac{1}{g(-\lambda)}\right)(1-p) = \beta (1-p),
\end{equation}

\noindent and it does not depend on $m$.

The values of $\alpha$ and $\beta$ can be used to estimate the characteristic times. Note that, at $\lambda = 0$ $\beta$ becomes 0. As in the case 
without sharing, at this value of $\lambda$ the dynamics is non-reversible, as seen from (\ref{m-update2-app}). As in the previous cases, the limit of a 
lethal phenotype~2 ($m \to \infty$) corresponds to $\alpha \to \infty$, so the invasion of a viable phenotype into a population of lethal phenotypes happens 
effortlessly.

Again, if the screening uses a different threshold value $x_0$, $x_2 < x_0 < x_1$, than the same formulas can be used with a different $m$: $m = \lceil 
x_0/x_2 \rceil$.

\section{General explicit update equations for quadratic and cubic replication functions}
\label{appE}
The computation of explicit form of equation (17) for the replication function $f(x) = a x^2$ with sharing, using the general 
algorithm for polynomial $f$ outlined in our aforementioned preprint (see \cite{Zadorin:2017}), gives the following result

\begin{equation}
\rho' = \frac{g_0 x^2 + g_1\left(2 x \bar x + \overline{x^2}\right) + g_2 \bar x^2}{\overline{x^2} + \lambda \bar x^2}\rho.
\end{equation}

\noindent Here all the averages are taken in respect to the distribution $\rho$ and

\begin{equation}
g_n \deq e^{-\lambda}\lambda^n\frac{d^n}{d\lambda^n}\left(\frac{e^\lambda - 1}{\lambda}\right).
\label{g-functions-app}
\end{equation}

One can show that $g_n$ can be defined recursively as

\begin{equation}
g_0 = g(\lambda),\quad g_n = \lambda^{n-1} - n g_{n-1},
\end{equation}

\noindent or explicitly as

\begin{equation}
g_n = (-1)^n n! \frac{1 - e^{-\lambda}}{\lambda} + \sum_{k=1}^n (-1)^{n-k} \frac{n!}{k!} \lambda^{k-1}.
\end{equation}

This allows to compute the new mean value of $x$ after one cycle of selection

\begin{equation}
\bar x' = \frac{g_0 \overline{x^3} + 3g_1 \bar x \overline{x^2} + g_2 \bar x^3}{\overline{x^2} + \lambda \bar x^2}.
\end{equation}

\noindent Here $\bar x'$ is the mean of $x$ computed with respect to the distribution $\rho'$. The difference of the mean activity is then computed as 
$\Delta \bar x = \bar x' - \bar x$.

Likewise, for the cubic replication function $f(x) = a x^3$ with sharing we obtain

\begin{equation}
\rho' = \frac{g_0 x^3 + g_1 \left(3(x^2 \bar x + x \overline{x^2}) + \overline{x^3}\right) +
3 g_2 \left(x \bar x^2 + \bar x \overline{x^2}\right) + g_3 \bar x^3}
{\overline{x^3} + 3 \lambda \bar x \overline{x^2} + \lambda^2 \bar x^3}\rho.
\end{equation}

\noindent and

\begin{equation}
\bar x' = \frac{g_0 \overline{x^4} + g_1 \left(4\bar x \overline{x^3} + 3 \overline{x^2}^2\right) + 6 g_2 \bar x^2 \overline{x^2} + g_3 \bar x^4}
{\overline{x^3} + 3 \lambda \bar x \overline{x^2} + \lambda^2 \bar x^3},
\end{equation}

The expressions for the case of a population of only two phenotypes, $x_1$ at frequency $p$ and $x_2$ at frequency $1-p$, can be easily derived from 
these general expressions using the probability density

\begin{equation}
\rho(x) = p\delta(x - x_1) + (1-p)\delta(x-x_2).
\end{equation}

\section{Asymptotics of $\Delta \bar x$ for a polynomial replication function at large $\lambda$ (additive activity)}
\label{appF}
\subsection{Preliminary notes}

In this section, we will heavily rely on the theory developed in our preprint \cite{Zadorin:2017}. We will use Schwartz's distributions to model 
probability density functions. The associated notations are not conventional in the field of applied probability theory, mathematical biology, and 
theoretical physics. Therefore, we will provide here a glossary that relates these general notations to more conventional (but less general/strict) 
equivalents. These equivalents can be used directly for well behaving probability density functions $\rho$ of activity distributions, for which they 
actually make sense.

\begin{eqnarray}
&\langle \rho,\phi\rangle \Leftrightarrow \intl{}{}\rho(x)\phi(x)\,dx, & \text{(application of }\rho\text{ to }\phi\text{)}\\
&\rho*\rho\,(x) \Leftrightarrow \intl{}{}\rho(y)\rho(x-y)\,dy,\\
&\delta_{x_1}(x) \Leftrightarrow \delta(x-x_1),\\
&\delta_x * \rho\,(y) \Leftrightarrow \rho(y-x).
\end{eqnarray}

We will also need the asymptotics of the functions $g_n$ defined in (\ref{g-functions-app})

\begin{equation}
g_n = \lambda^{n-1} - n\lambda^{n-2} + o(\lambda^{n-2}),\quad \lambda \to \infty.
\label{asymptotics-app}
\end{equation}

\subsection{Power functions $f(x) = a x^m$}

We have shown in the preprint \cite{Zadorin:2017} that the update equation for the probability density of the activity distribution in the case of the 
additive activity, with replication function $f$, and with offspring sharing, is given by the formula

\begin{equation}
\rho' =
\frac{\displaystyle \sum_{n=0}^\infty P_n (s^m_n)_x}{\displaystyle \sum_{n=0}^\infty P_n s^m_n}\,\rho =
\frac{S^m_x}{S^m}\,\rho,
\end{equation}

\noindent where

\begin{equation}
P_n = \frac{e^{-\lambda}\lambda^n}{(n+1)!},\quad (s^m_n)_x = \langle\delta_x*\rho^{*n},y^m\rangle, \quad
s^m_n = \langle\rho^{*n+1},y^m\rangle, \quad S^m_x = \sum_{n=0}^\infty P_n (s^m_n)_x, \quad S^m = \sum_{n=0}^\infty P_n s^m_n.
\end{equation}

It follows that the mean of the distribution is updated according to

\begin{equation}
\bar x' = \frac{\langle S^m_x \rho, x\rangle}{S^m}.
\label{mean-update-app}
\end{equation}

We will be interested only by the two highest order, in $\lambda$, terms in both $S^m_x$ and in $S^m$.

The interaction of the multiplication by the argument and convolution products of distributions $y(\rho_1*\ldots*\rho_k)$ is equivalent to the interaction of the differentiation by the parameter and the product of functions $\D (\psi_1\ldots\psi_k)$. This analogy becomes exact upon the application of the Laplace transform. In particular, the following Leibniz rule holds: $y(\rho_1*\rho_2) = (y\rho_1)*\rho_2 + \rho_1*(y\rho_2)$.

Note that

\begin{multline}
\D^m(\phi\psi^n) = A^m_n\phi\psi^{n-m}(\D\psi)^m +\\+
m A^{m-1}_n(\D\phi)\psi^{n-m+1}(\D\psi)^{m-1}+\frac{m(m-1)}{2} A^{m-1}_n\phi\psi^{n-m+1}(\D\psi)^{m-2}\D^2\psi + \ldots
\label{expansion-functions-app}
\end{multline}

\noindent where $A^k_n \deq n(n-1)\ldots(n-k+1)$, $A^0_n \deq 1$, the ommited terms have the form $p(m)A^k_n\Pi$, where $p(m)$ is a polynomial that does not depend on $n$, $\Pi$ is a product of functions $\phi$, $\psi$ and their derivatives, and $k < m-1$. Indeed, the multipliers $A^k_n$ appear only as a result of the derivation of $\psi^n$ $k$ times. Therefore, for every $m$, the term with $A^m_n$ (the term in the first line of (\ref{expansion-functions-app})) is always unique and comes only form the $m$-th differentiation of $\psi^n$. The terms with $A^{m-1}_n$ come from two sources: from the application of $\D$ to this term obtained for $\D^{m-1}$, when either $\phi$ or $(\D \psi)^{m-1}$ is differentiated once, and from the application of $\D$ to the power of $\psi$ in the terms of the second line for $m-1$. Thus, the coefficient in front of $A^{m-1}_n$ in the first term of the second line in (\ref{expansion-functions-app}) increases by $1$ with the increase of $m$, while this coefficient of the second term in the same line increases by $m$ with the increase of $m$. This information is enough to recover the exact expression for these terms that are shown in (\ref{expansion-functions-app}).

As a consequence, we have

\begin{multline}
y^m(\delta_x*\rho^{*n}) = A^m_n\,\delta_x*\rho^{*n-m}*(y\rho)^{*m} +\\
+ m A^{m-1}_n (y\delta_x)*\rho^{*n-m+1}*(y\rho)^{*m-1}
+ \frac{m(m-1)}{2} A^{m-1}_n \delta_x*\rho^{*n-m+1}*(y\rho)^{*m-2}*(y^2\rho) + \ldots
\end{multline}

\noindent We retain only these terms, as $\sum\limits_{n=0}^\infty P_n A^k_n = g_k$ and we are interested only in the two highest order terms in $\lambda$. After summation with $P_n$, taking into account $\langle y\delta_x,1\rangle = x$, $\langle x\rho,1\rangle = \bar x$, $\langle x^2\rho,1\rangle = \overline{x^2}$, $\langle \rho_1*\rho_2,1\rangle = \langle \rho_1,1\rangle \langle \rho_2,1\rangle$, we have

\begin{equation}
S^m_x = g_m \bar x^m + m g_{m-1} x \bar x^{m-1} + \frac{m(m-1)}{2} g_{m-1} \bar x^{m-2} \overline{x^2} + \ldots,
\end{equation}

\noindent and, as a consequence, with the asymptotics (\ref{asymptotics-app})

\begin{equation}
\langle S^m_x \rho, x\rangle = g_m \bar x^{m+1} + \frac{m(m+1)}{2} g_{m-1} \bar x^{m-1}\overline{x^2} + \ldots =
\lambda^{m-2} \bar x^{m-1}\left(\lambda \bar x^2 - m \bar x^2 + \frac{m(m+1)}{2} \overline{x^2}\right) + o(\lambda^{m-2}),\; \lambda \to \infty.
\end{equation}

In the same way we have

\begin{equation}
y^m \rho^{*n+1} = A^m_{n+1} \rho^{*n-m+1} * (y\rho)^m + \frac{m(m-1)}{2} A^{m-1}_{n+1} \rho^{*n-m+2} * (y\rho)^{*m-2} *(y^2\rho) + \ldots
\end{equation}

\noindent and, using $\sum\limits_{n=0}^\infty P_n A^k_{n+1} = \lambda^{k-1}$ for $k > 0$,

\begin{equation}
S^m = \lambda^{m-2}\bar x^{m-2} \left(\lambda \bar x^2 + \frac{m(m-1)}{2} \overline{x^2}\right) + o(\lambda^{m-2}),\; \lambda \to \infty.
\end{equation}

Finally, the change of the mean in one cycle of selection is then given by

\begin{equation}
\Delta x = \bar x' - \bar x = \frac{\lambda^{m-2}\bar x^{m-1}\left(\lambda \bar x^2 - m\bar x^2 + \frac{m(m+1)}{2}\overline{x^2}\right) + 
o(\lambda^{m-2})}
{\lambda^{m-2}\bar x^{m-2}\left(\lambda \bar x^2 + \frac{m(m-1)}{2}\overline{x^2}\right) + o(\lambda^{m-2})} - \bar x,\quad \lambda \to \infty,
\end{equation}

\noindent or after simplification

\begin{equation}
\Delta \bar x = \frac{m}{\lambda}\,\frac{\overline{x^2} - \bar x^2}{\bar x} + o(1/\lambda),\quad \lambda \to \infty.
\end{equation}

Note that this equation is formally correct even for $m = 0$.

\subsection{General polynomial functions $f(x) = a_m x^m + \ldots + a_1 x + a_0$}

When $f$ has the form $f(x) = a_m x^m + \ldots + a_1 x + a_0$, $a_m \neq 0$, equation (\ref{mean-update-app}) is rewritten as

\begin{equation}
\bar x' = \frac{a_m \langle \rho, xS^m_x\rangle + \ldots + 
a_1 \langle \rho, xS^1_x\rangle + a_0}
{a_m S^m + \ldots + 
a_1 S^1 + a_0}.
\end{equation}

As we have seen before, the contribution to the two highest order in $\lambda$ are given only by the term with $a_m$ and the term with $a_{m-1}$. More 
specifically, the asymptotics of $\Delta \bar x$ takes the form

\begin{equation}
\Delta x = \bar x' - \bar x = \frac{\lambda^{m-2}\bar x^{m-1}\left(a_m\lambda \bar x^2 - a_m m\bar x^2 + a_m \frac{m(m+1)}{2}\overline{x^2}
+ a_{m-1} \bar x\right) + 
o(\lambda^{m-2})}
{\lambda^{m-2}\bar x^{m-2}\left(a_m \lambda \bar x^2 + a_m \frac{m(m-1)}{2}\overline{x^2}
+ a_{m-1} \bar x\right) + o(\lambda^{m-2})} - \bar x,\quad \lambda \to \infty,
\end{equation}

It is not difficult to see that the effect of the terms with $a_{m-1}$ cancels with $\lambda \to \infty$ in the leading order. Finally, we obtain the 
following asymptotics

\begin{equation}
\Delta \bar x = \frac{m}{\lambda}\,\frac{\overline{x^2} - \bar x^2}{\bar x} + o(1/\lambda),\quad \lambda \to \infty.
\end{equation}

\section{Selection with limited resources, theoretical considerations}
\label{appG}
Figure~6B shows the results of numerical experiments with discrete time. However, we can anticipate its result using the continuous 
time approximation. We have already seen in Appendix~\ref{appA}, Section~2 that the effect of $\lambda$ on the selection dynamics is a mere time rescaling by 
the scaling factor $g(\lambda)$. This stays true for general phenotypical distributions $\rho$ \cite{Zadorin:2017}. Let the fixed droplet 
budget be $M_\mathrm{tot}$, the number of rounds be $t$, and the fixed population size at each step be $N$. Then the number of compartments per round 
is equal to $M = M_\mathrm{tot}/t$ and the mean occupancy is equal to $\lambda = N/M = t N/M_\mathrm{tot} = at$, where $a = N/M_\mathrm{tot} = 
\mathrm{const}$. The actual progress of the selection (the reached valued of $p$) after $t$ rounds is expected to be the same as for the case $\lambda 
= 0$ at time $g(\lambda) t$ (for the same initial $p$). But $g(\lambda)t = g(at)at/a$, and $g(x)x = 1 - e^{-x}$ is an increasing function of $x$. 
Therefore, the larger $\lambda$ at each round (the more overall rounds is used) the better the overall selection progress. As this functions is also 
bounded from above, it has the upper bound: $\sup\limits_x g(x)x = 1$. If we again measure the selection progress in the time $t_e$ it takes the case 
$\lambda = 0$ to reach the same value of $p$ from the same initial state, then this equivalent time has the upper bound $\bar t_e = t/a = t 
M_\mathrm{tot}/N$. The value of $p$ achieved by the system with $\lambda = 0$ by this time $\bar t_e$ from the initial frequency sets the maximal 
theoretical gain from the budget management. This upper bound explains the saturation effect with the increased number of rounds seen on Figure~6B.

\section{Details on the $N\!K$-landscape used to generate Figure 6A}
\label{appH}
\begin{figure}[thb]
\includegraphics[scale=0.5]{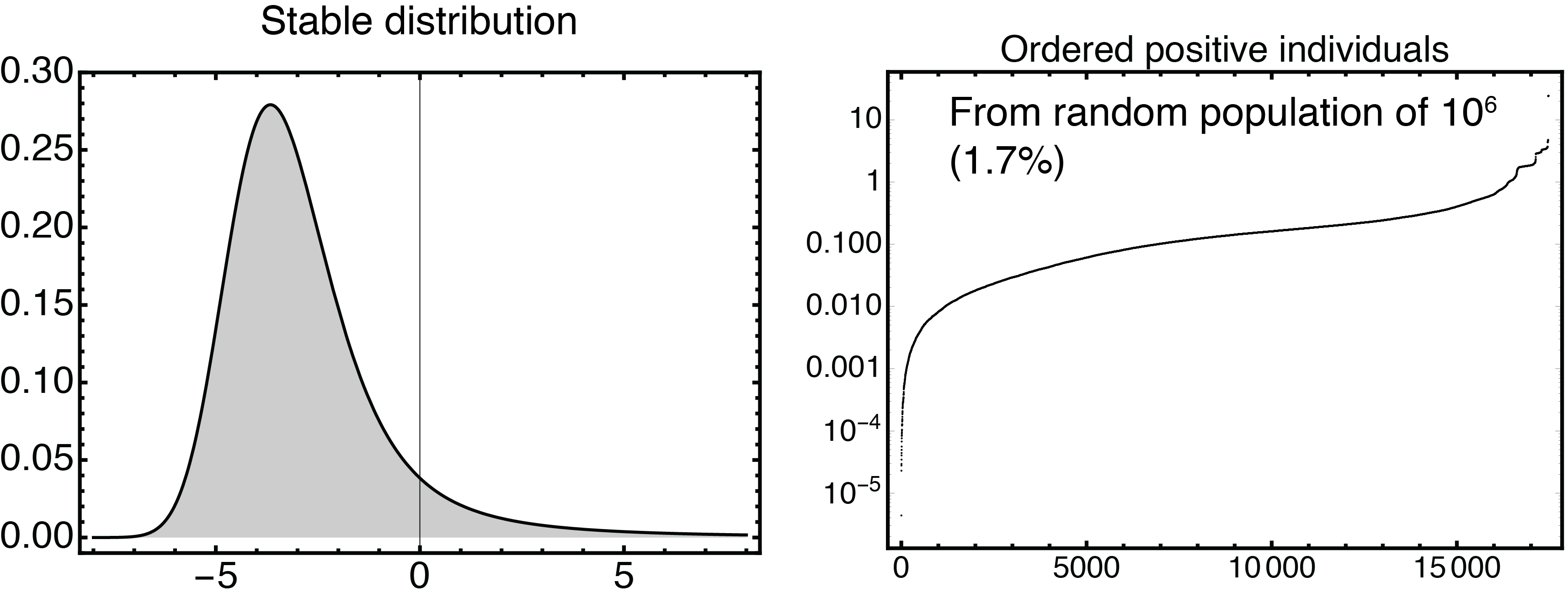}

\caption{Statistics of the $N\!K$-landscape used in the study. Left panel: the stable distribution of activity contributions. Right panel: the rank 
ordering of nonlethal individuals by the activity in a generated population of size $10^6$. Note the presence of a large plateau of low activity with 
a tiny portion of high activity mutants.}

\label{nk}
\end{figure}

The $N\!K$ fitness landscape was shaped to approximate some assumed characteristics of enzymatic landscapes. This include an extended network of epistatic 
interactions (relatively high value of $K$ have been reported, \cite{Kauffman:1989nk}), the presence of many non-functional variants, and a large 
dynamic range of activity values. The values for the lookup table of activity contributions are drawn from a stable distribution with index of 
stability 1.5, skewness parameter 1, centered at $-2.5$ (see Fig.~\ref{nk}). Moreover, activity of an individual is set to 0 if the sum of contributions 
is negative. This leads to a fraction of non-lethal variants of 1.7\%.


\bibliography{bibliography}

\end{document}